\documentclass{aa} 
\usepackage{times,amssymb,latexsym,amsmath} 
\usepackage{epsfig} 

\title{Deep Ly$\alpha$ imaging of two z=2.04 GRB host galaxy fields
\thanks{Based on observations made with the Nordic Optical Telescope,
operated on the island of La Palma jointly by Denmark, Finland,
Iceland, Norway, and Sweden.}\fnmsep\thanks{Based on observations made
with the NASA/ESA Hubble Space Telescope, obtained from the data archive
at the Space Telescope Institute. STScI is operated by
the association of Universities for Research in Astronomy, Inc.
under the NASA contract  NAS 5-26555.}
}

\author{J.P.U. Fynbo \inst{1}
   \and P. M\o ller \inst{1} 
   \and B. Thomsen \inst{2}
   \and J. Hjorth \inst{3} 
   \and J. Gorosabel \inst{4,5,6}
   \and M.I. Andersen \inst{7}
   \and M.P. Egholm  \inst{8,2,1}
   \and S. Holland \inst{9}
   \and B.L. Jensen \inst{3}
   \and H. Pedersen \inst{3}
   \and M. Weidinger \inst{2,1}
   }

\institute{
      European Southern Observatory,  
      Karl-Schwarzschild-Stra\ss e 2,
      D--85748, Garching by M\"unchen, Germany
      \and
      Institute for Physics and Astronomy
      University of {\AA}rhus,
      Ny Munkegade, DK--8000 {\AA}rhus C, Denmark
      \and
      Astronomical Observatory,
      University of Copenhagen,
      Juliane Maries Vej 30, DK--2100 Copenhagen \O, Denmark
      \and
      Danish Space Research Institute,
      Juliane Maries Vej 30, DK--2100 Copenhagen \O, Denmark
      \and
      Laboratorio de Astrof\'{\i}sica Espacial y F\'{\i}sica
      Fundamental (LAEFF-INTA), P.O. Box 50727, E--28080 Madrid, Spain
      \and
      Instituto de Astrof\'{\i}sica de Andaluc\'{\i}a              
      (IAA-CSIC), P.O. Box 03004, E--18080 Granada, Spain
      \and
      Division of Astronomy,
      P.O. Box 3000, FIN--90014 University of Oulu,
      Finland
      \and
      Nordic Optical Telescope, Apartado Postal 474,
      38700 Santa Cruz de La Palma, Canary Islands, Spain
      \and
      Department of Physics, University of Notre Dame,
      Notre Dame, IN 46556-5670, U.S.A.
      }
\mail{jfynbo@eso.org}
\offprints{J.P.U. Fynbo}
\mail{jfynbo@eso.org}

\date{Received  / Accepted }

\begin{document}
\titlerunning{Deep Ly$\alpha$ imaging of two z=2.04 GRB host galaxy
fields}

\abstract{
We report on the results of deep narrow-band Ly$\alpha$ and broad-band U
and I imaging of the fields 
of two Gamma-Ray bursts at redshift z=2.04 (GRB~000301C and GRB~000926). 
We find that the host galaxy of GRB~000926 is an extended (more 
than 2 arcsec), strong Ly$\alpha$ emitter with a rest-frame 
equivalent width of 71$^{+20}_{-15}$ \AA. The galaxy consists of two
main components and several fainter knots. GRB~000926 occurred in the
western component, whereas most of the Ly$\alpha$ luminosity
(about 65\%) originates in the eastern component. Using archival HST 
images of the host galaxy we measure the spectral slopes 
($f_{\lambda} \propto \lambda^{\beta}$) of the two
components to $\beta$ = $-$2.4$\pm$0.3 (east) and $-$1.4$\pm$0.2 (west).
This implies that both components contain at most small amounts of
dust, consistent with the observed strong Ly$\alpha$ emission.
The western component has a slightly redder V$-$I colour than the
eastern component, suggesting the presence of at least some dust.
We do not detect the host galaxy of GRB~000301C in neither 
Ly$\alpha$ emission nor in U and I broad-band images. The strongest 
limit comes from combining the narrow and U-band imaging where we infer 
a limit of U(AB)$>$27.7 (2$\sigma$ limit per arcsec$^2$). The upper
limits on the Ly$\alpha$ flux implies a Ly$\alpha$ equivalent width
upper limit of $\sim 150$ \AA.
We find 
eleven and eight other galaxies with excess emission in the narrow filter
in the fields of GRB~000301C and GRB~000926 respectively. These galaxies 
are candidate Ly$\alpha$ emitting galaxies in the environment of 
the host galaxies. Based on these detections we conclude that GRB~000926 
occurred in one of the strongest centres of star formation within several 
Mpc, whereas GRB~000301C occurred in an intrinsically very faint galaxy far 
from being the strongest centre of star formation in its galactic 
environment. Under the hypothesis that GRBs trace star formation,
the wide range of GRB host galaxy luminosities implies a very steep
faint end slope of the high redshift galaxy luminosity function.
\keywords{cosmology: observations --
gamma rays: bursts}
}

\maketitle

\section{Introduction}
In addition to the very important question of what may be
the central engine of Gamma-Ray Bursts (GRBs, see van Paradijs et al. 2000
for a recent review) these enigmatic
events constitute a powerful new tool for the study of the intermediate
and high redshift galaxy population. This is mainly because GRB selection
of galaxies is independent of other selection methods such as
continuum flux (e.g. Steidel \& Hamilton 1992; Steidel et al. 1996; 
Adelberger \& Steidel 2000;
Fontana et al. 2000), gas absorption (Wolfe et al. 1986; 
M\o ller \& Warren 1993, 1998; Leibundgut \& Robertson 1998; Fynbo et
al. 1999; Kulkarni et al. 2000, 2001; Ellison et al. 2001; 
Warren et al. 2001; M\o ller et al. 2001) or Ly-$\alpha$ emission (see
below). By comparing the results of these very different methods by which 
to select high redshift galaxies we can better understand the 
selection biases inherent to each method and hence obtain a more
complete understanding of the underlying population of high redshift
galaxies (see also M\o ller et al. 2001). 

Since the first detection of 
an optical afterglow of a GRB in 1997 (van Paradijs et al. 
1997) the detection of about 25 GRB host galaxies have been reported
(Hogg \& Fruchter 
1999 and references therein; Bloom et al. 1999; Holland \& Hjorth 
1999; Hjorth et al. 2000; Klose et al. 2000; Fruchter 2001; Vreeswijk 2001; 
Castro-Tirado et al. 2001; Holland et al. 2001; Fynbo et al. 2001a, 
Fruchter et al. 2001a; Kaplan et al. 2001; Fruchter et al. 2001b; Bloom
et al. 2001). Only for GRB~990308 has an unambiguous 
detection of the host galaxy not yet been reported.
GRB host galaxies have been detected in a very broad range of 
redshifts from z=0.43 (GRB~990712) to z=4.50 (GRB~000131) excluding GRB~980425 at z=0.0085
which may be a burst of different nature. Furthermore, the measured
optical magnitudes of GRB hosts also span a very wide range from V=22.8
(GRB~980703) to V$\approx$30 (GRB~980326, GRB~990510, GRB~000301C).
This span is both due to a wide range of distance moduli and
a large spread in absolute magnitudes.
In fact, GRB-selection of galaxies is currently the only technique that 
allows the detection of galaxies over such a wide range of redshifts
and absolute luminositites.
Most host galaxies detected so far are estimated to be fainter than 
L$^*$ at similar redshifts (L$^*$ is here defined from the fit to a 
Schechter function to the observed luminosity function at the given
redshift). This fact may be due to intrinsic extinction 
in the GRB host galaxies (Sokolov et al. 2001), but it is not known if GRB 
host galaxies have more than average intrinsic extinction compared to
other galaxies at similar redshifts. In general GRB host galaxies are 
actively star-forming, i.e. 
characterised by restframe UV emission and large emission-line 
equivalent widths (Bloom et al. 2002; Fruchter et al. in prep.). 
Searches for sub-mm emission from GRBs (Smith et al. 1999, 2001) show that 
GRB host galaxies are in general not similar to the obscured star bursts 
detected in deep SCUBA and ISO surveys (e.g.  Ivison et al. 2000; 
Franceschini et al. 2001).

In this paper we present a study of the host galaxies and environments
of the two z$\approx$2 bursts GRB~000301C and GRB~000926. The goal 
is to place the host galaxies and their galactic environments
in the context of other z$\approx$2 galaxies. The redshifts
of the two bursts have both been determined via absorption lines in the
spectra of the optical afterglow. GRB~000301C was at z=2.040 (Smette et al. 
2001; Castro et al. 2000; Jensen et al. 2001) and GRB~000926 
at z=2.038 (Fynbo et al. 2001c; Castro et al. 2001; M{\o}ller et al. 
in prep.). The host galaxy of 
GRB~000301C is very faint with estimates of its magnitude between
R=28.0 (Bloom et al. 2002) and R=29.7 (Levan, private communication; 
Fruchter et al. in prep.), whereas in the case of GRB~000926 Fynbo et 
al. (2001a) and Price et al. (2001) report the detection of an extended 
R=24 galaxy consisting of several compact knots near the position of the 
optical 
afterglow. Here we use narrow-band Ly$\alpha$ observations to probe these
two host galaxies and to search for other galaxies in their 
environments. It is presently unknown whether the galactic environments 
of GRB host galaxies are overdense or representative of the general
field. Narrow-band Ly$\alpha$ observations are ideal to address this
question as it has been shown to be an efficient way to probe the faint
end of the galaxy population at redshifts 2$<$z$<$5 (M\o ller \& Warren
1993; Francis et al. 1995; Pascarelle et al. 1996; Thommes et al. 1997;
Pascarelle et al. 1998; Cowie \& Hu 1998; Hu et al. 1998; Fynbo et al.
1999, 2000; Kudritzki et al. 2000; Kurk et al. 2000; Pentericci et al.
2000, Steidel et al. 2000; Roche et al. 2000; Rhoads et al. 2000; Malhotra
\& Rhoads 2002; Fynbo et al. 2001b). Ly$\alpha$ selection has 
the advantage that it is not continuum flux limited and hence allows
the detection of high redshift proto--galaxies, or Ly$\alpha$ Emitting
Galaxy-building Objects (LEGOs,  M{\o}ller \& Fynbo 2001), 
that are intrinsically much fainter than those selected in continuum flux 
limited 
surveys. However, since not all galaxies are Ly$\alpha$ emitters (about
30\% of the R$<$25 z$\approx$3 Lyman-Break galaxies are (Steidel et al.
2000)), galaxy
samples selected from Ly$\alpha$ emission will not be complete to the
continuum flux limit of the sample.

The paper is organised in the following way. In Sect.~\ref{obs} we
describe the observations and the data reduction, in
Sect.~\ref{results} we present our results and in Sect.~\ref{discuss}
we discuss the implications of our results. 
We have assumed a cosmology where $H_0$ = 65 km s$^{-1}$ Mpc$^{-1}$, 
$\Omega_m = 0.3$, and $\Omega_\Lambda = 0.7$. For this cosmology a 
redshift of 2.04 corresponds to a luminosity distance 
$d_{\rm{lum}}$ = 17.16 Gpc and a distance modulus of 46.17.  One arcsecond 
corresponds to 9.00 proper kpc and the look-back time is 11.1 Gyr.

\section{Observations and Data reduction}
\label{obs}

The observations were carried out during five dark, photometric/clear 
nights in May 2001 and four dark nights in August 2001 at the 
2.56-m Nordic Optical Telescope (NOT)
using the Andaluc\'{\i}a Faint Object Spectrograph  and Camera (ALFOSC). The 
ALFOSC detector is a 2048$^2$ pixels thinned Loral CCD with a pixel
scale of 0\farcs189.

\begin{figure}
\begin{center}
\epsfig{file=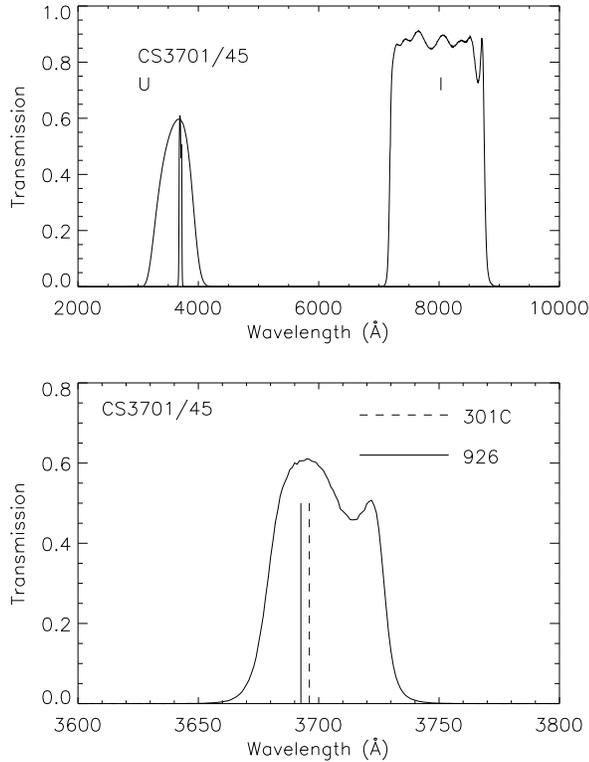,width=8cm}
\caption{{\it Upper panel:} The transmission curves of the three
filters used in this study, namely the narrow CS3701/45 filter
and the broad-band U and I filters. {\it Lower panel:} The transmission
curve of the CS3701/45 filter. The vertical lines mark the wavelengths
of Ly$\alpha$ redshifted to
GRB~000301C (dashed line) and GRB~000926 (full-drawn line).}
\label{filtercurves}
\end{center}
\end{figure}

The fields of GRB~000301C and GRB~000926 were imaged
in three filters: the standard U and I filters and a special
narrow-band filter manufactured by Custom Scientific Inc. The 
narrow-band filter (CS3701/45) is tuned to Ly$\alpha$ at z=2.04 
and has a width of 45 \AA \ (corresponding to a redshift width of
$\Delta z = 0.037$ for Ly$\alpha$ or a Hubble flow depth of 3600 km
s$^{-1}$). The filter transmission curves of the
CS3701/45, U and I filters are shown in Fig.~\ref{filtercurves}.

In order to reduce the contribution to the total noise from 
read-out noise, the detector was binned 2 by 2 during the
narrow-band observations, and the integration times for the
individual exposures was set to 4000 s. 
The total integration times for the GRB~000301C field was 
12.2 hours (CS3701/45), 7.3 hours (U-band), and 3.0 hours 
(I-band). For the field of GRB~000926 total integration times
were 12.2 hours (CS3701/45), 7.3 hours (U-band), and 3.1 hours 
(I-band). The journal of observations is presented in 
Table~\ref{obs-journal}.
The individual exposures were bias-subtracted and flat-field
corrected using standard techniques. The individual reduced
images were combined using a $\sigma$-clipping routine
optimised for faint sources (for details see M{\o}ller \& Warren, 
1993). The full-width-at-half-maximum (fwhm) of point sources in 
the combined images are 0\farcs83 (I-band), 1\farcs25 (U-band) and 
1\farcs24 (CS3701/45). 

%=====================Begin Table 1==============================
\begin{table}
\begin{center}
\caption{The log of observations at the NOT, May and August 2001.}
\begin{tabular}{@{}lllcccc}
\hline
date        & filter & Exp.time 301C & Exp.time 926 \\
            &        &     (s)    &    (s)     \\
\hline
May 20      & CS3701/45 & 8000      &   8000       \\
May 20      & U         & 2500      &   3750       \\
May 20      & I         & 3200      &   2400       \\  
May 21      & CS3701/45 & 8000      &   8000       \\
May 21      & U         & 2500      &   2500       \\
May 21      & I         & 2100      &   3300       \\  
May 22      & CS3701/45 & 11800     &  11800       \\
May 22      & U         &  -        &   1250       \\
May 22      & I         & 2700      &   1700       \\  
May 23      & CS3701/45 & 8000      &   8000       \\
May 23      & U         & 2500      &   1250       \\
May 23      & I         & 1200      &   1500       \\  
May 24      & CS3701/45 & 8000      &   8000       \\
May 24      & U         & 3750      &   2500       \\
May 24      & I         & 1500      &   2100       \\  
Aug 13      & U         &  -        &   3000       \\
Aug 14      & U         & 3000      &   4500       \\
Aug 15      & U         & 6000      &   3000       \\
Aug 16      & U         & 6000      &   4500       \\
\hline
\label{obs-journal}
\end{tabular}
\end{center}
\end{table}
%=====================End Table 1===============================

In order to calibrate the narrow-band observations the four HST 
spectrophotometric standard stars feige66, feige67, BD+25$^o$4655, 
and BD+33$^o$2642 (Colina \& Bohlin 1994) were observed. We 
obtained well sampled Spectral Energy Distributions (SEDs) of
these four stars from the archive of the HST Calibration Data Base 
System\footnote{ftp://ftp.stsci.edu/cdbs/cdbs1/calspec/}. We then
calculated the AB magnitudes of the stars in the narrow filter
as the weighted mean of the AB magnitudes within the filter, where 
the filter transmission was used as weight. We find AB magnitudes 
of 9.96, 11.08, 9.06,
and 10.64 for feige66, feige67, BD+25$^o$4655, and BD+33$^o$2642 
respectively. Instrumental magnitudes were measured in a 11 arcsec
circular aperture. To the difference between standard 
AB magnitudes and instrumental magnitudes as a function of airmass
we finally fitted a zero-point and an extinction term. For counts 
given as electrons per second we find a zero-point of 21.57 and an 
extinction term of 0.42 magnitudes per unit airmass. 
The broad-band images were calibrated using  secondary standards 
from Henden et al. (2000) and brought onto the AB-system using the
transformations given in Fukugita et al. (1995): U(AB) = U+0.69;
I(AB) = I+0.43.

\section{Results}
\label{results}

\begin{figure*}
\begin{center}
\epsfig{file=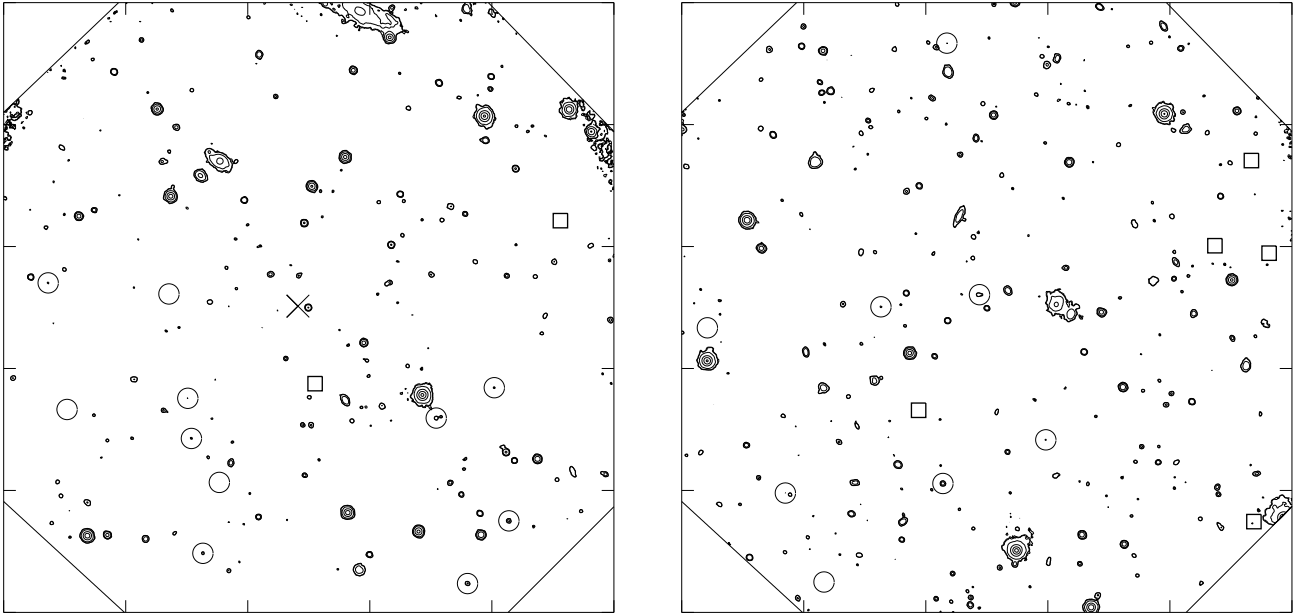, width=18cm, clip=}
\caption{The combined images of the field of GRB~000301C ({\it left}) 
and GRB~000926 ({\it right}) each based on 12.2 hours of narrow-band imaging.
The positions of 19 candidate emission-line galaxies in the two fields have 
been 
marked with circles. The position of the host galaxy of GRB~000301C, for which 
no emission is detected, is marked with ``$\times$''. The positions of 7
galaxies with a deficit of flux in the narrow-band filter are shown with
squares.  The corners of the images are lost 
due to vignetting. The effective area of each field is 27.6 arcmin$^2$. 
East is left and north is up. The distance between the tick-marks is
1.14 arcmin.}
\label{fields}
\end{center}
\end{figure*}

Contour plots of the combined narrow-band images of the 340$\times$340
arcsec$^2$ fields surrounding the two GRB positions are shown in 
Fig.~\ref{fields}. The position of 19 candidate LEGOs (see 
Sect.~\ref{lego} below) are indicated with circles, and the position of
7 galaxies with a deficit of flux in the narrow filter (see
Sect.~\ref{abs}) are shown with squares.
The position of GRB~000301C is indicated with an `$\times$'.

\subsection{Object Detection}

In order to create a catalogue of reliable candidate emission-line
galaxies we need robust detection and precise colours. We use the
software SExtractor (Bertin \& Arnouts, 1998) for both detection and
photometry. SExtractor uses a detection image to build the
isophote apertures and then measures the fluxes of objects in the
narrow, U, and I images in these apertures.
Because the narrow band is inside the U-band, the optimal detection
of emission line objects is obtained on a detection image created as
a weighted sum of the combined narrow-band and U-band images. We found
that the weights which optimized the Signal-to-Noise (S/N) ratio of
emission line objects were 0.7 and 0.3 for the narrow-band and U-band
images respectively. This combined image was then convolved with a
Gaussian filter with a width similar to that of the combined seeing.
For the SExtractor input detection parameters we used a minimum area of
5 (binned) pixels 1.1$\sigma$ above the noise in the background. In our
final catalogue we only include objects detected at a total S/N ratio 
larger than 5 in an isophotal aperture in the narrow-band.
In total we found 648 objects in the two fields. Objects detected
at the 5$\sigma$ levels in the narrow-band have AB isophotal magnitudes
of approximately 25.5.\footnote{In the following we shall 
denote with ``n'' the magnitude in the narrow filter.}
This corresponds to an observed 
emission line flux of 2$\times$10$^{-17}$ erg s$^{-1}$ cm$^{-2}$
(for details on the photometry see Sect.~\ref{photometry}).
In a 3\farcs0 circular aperture the 5$\sigma$ limit is n(AB)=25.0 
corresponding to 3.5$\times$10$^{-17}$ erg s$^{-1}$ cm$^{-2}$.
The combined U-band image is 1 magnitude deeper for continuum sources,
i.e. U(AB)=26.5 at 5$\sigma$ significance in the isophotal aperture.

\subsection{Photometry}
\label{photometry}

To derive colours for all detected objects we measure the 
{\it isophotal} magnitudes. The isophotal apertures are defined on 
the detection image, and the same (small) aperture is then used for
all three frames narrow, U and I. To the extent that the seeing of 
the images in the different passbands are equal, which is the case 
for the U and narrow-band images, and if the morphologies of the 
objects are independent of colour the {\it isophotal} magnitudes 
therefore provide photometry appropriate for accurate colour 
determination. The systematic error on the U-I colours caused by the 
slightly better seeing in the combined I image is much smaller than the 
photometric errors and hence ignored.
To estimate the total magnitudes of the objects
the isophotal aperture is too small. Instead we use aperture photometry
with 3\farcs0 and 4\farcs0 apertures and determine aperture corrections
based on the radial profiles of the candidate emission galaxies as described 
in Sect.~\ref{extended}.

We calculate two-sided error-bars on magnitudes
as 
\[
\sigma(m)^+ = \rm{mag}(\rm{flux}-\sigma(flux))-mag(flux) 
\]
\vskip -0.7cm
\[\sigma(m)^- = \rm{mag}(\rm{flux}) - mag(flux+\sigma(flux)),
\]
\noindent
where $\sigma(flux)$ is derived by SExtractor.
We calculate error-bars on colours using the maximum
likelihood analysis described in appendix A.

\begin{figure}
\begin{center}
\epsfig{file=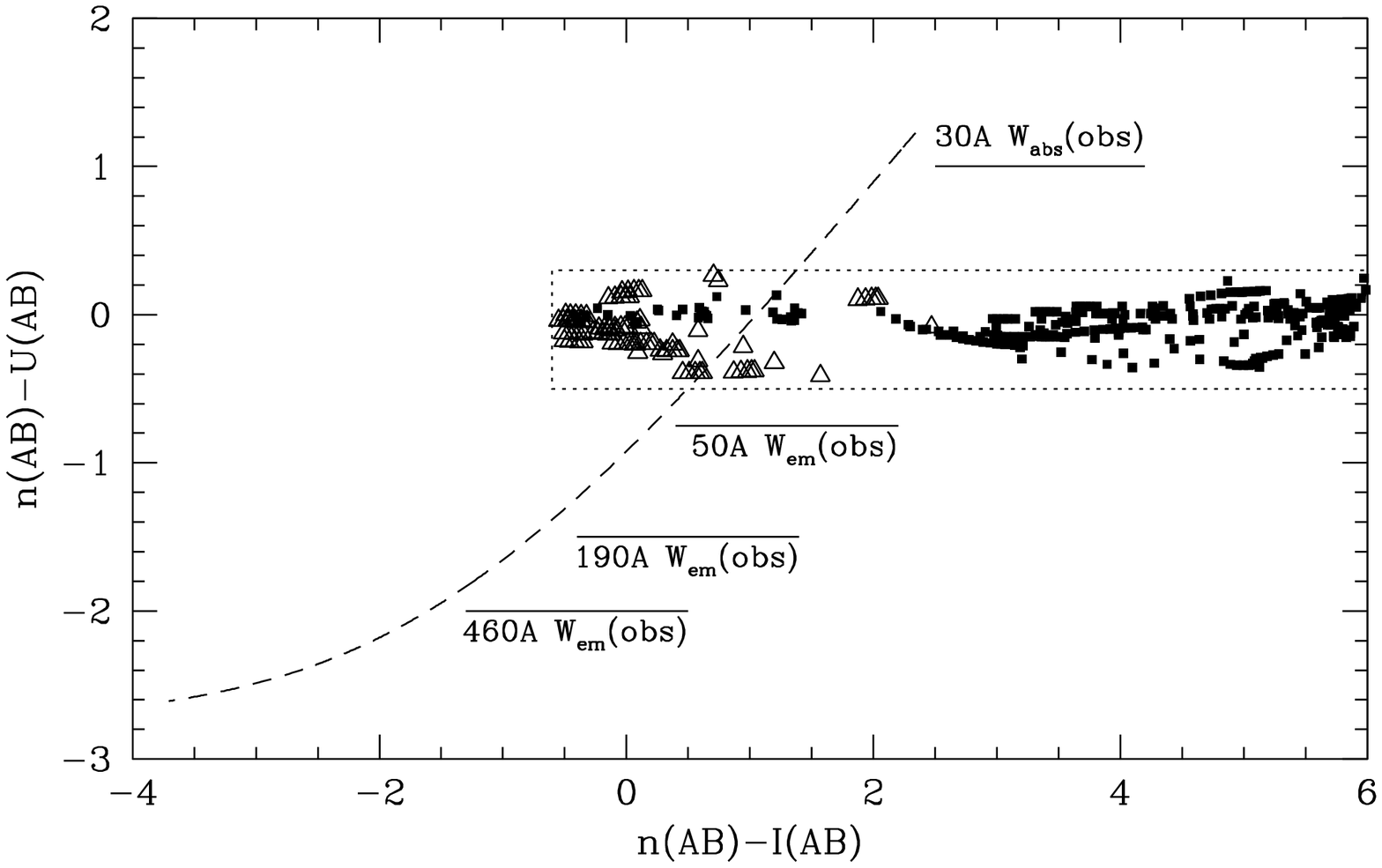, width=9cm, clip=}
\epsfig{file=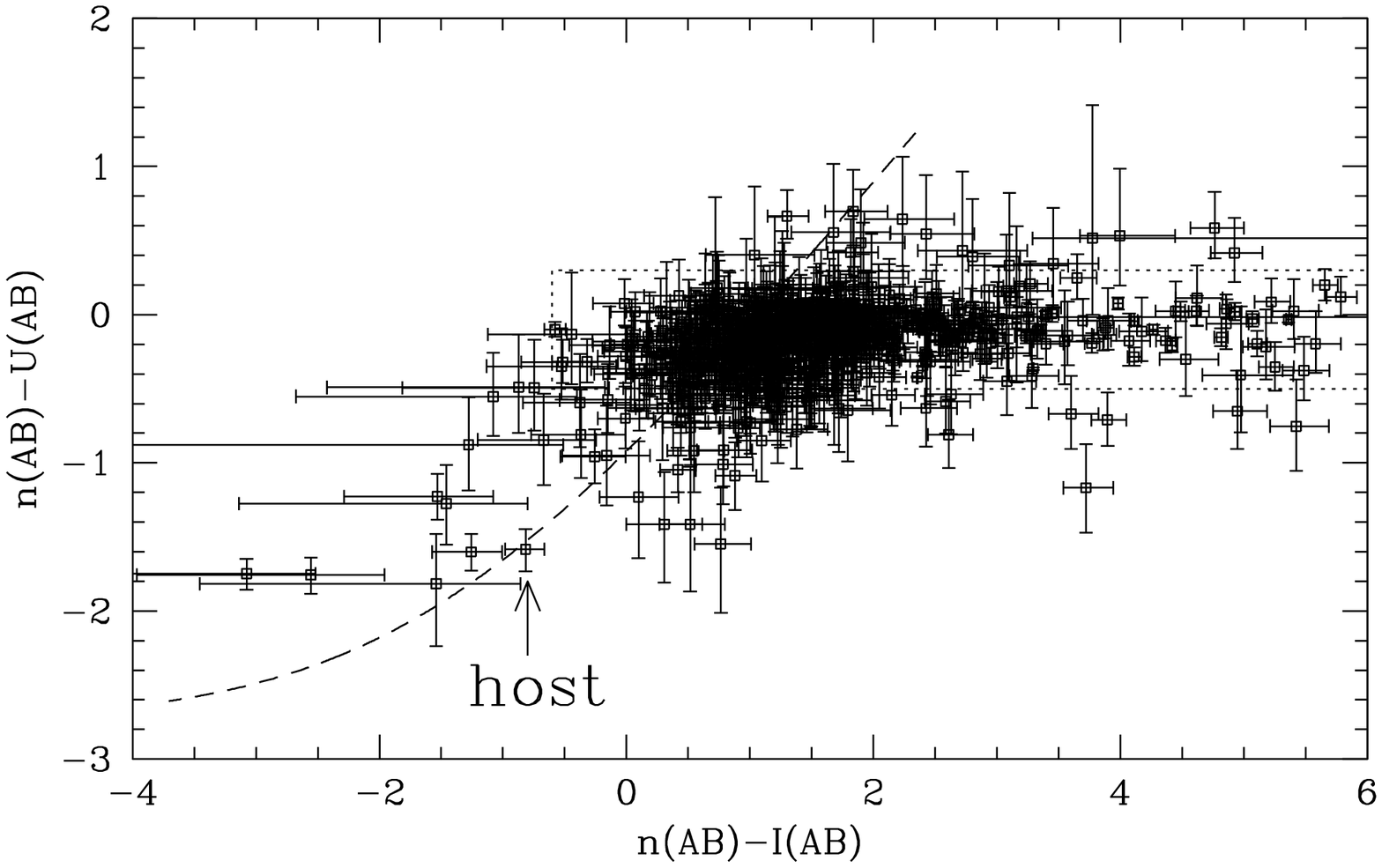, width=9cm, clip=}
\epsfig{file=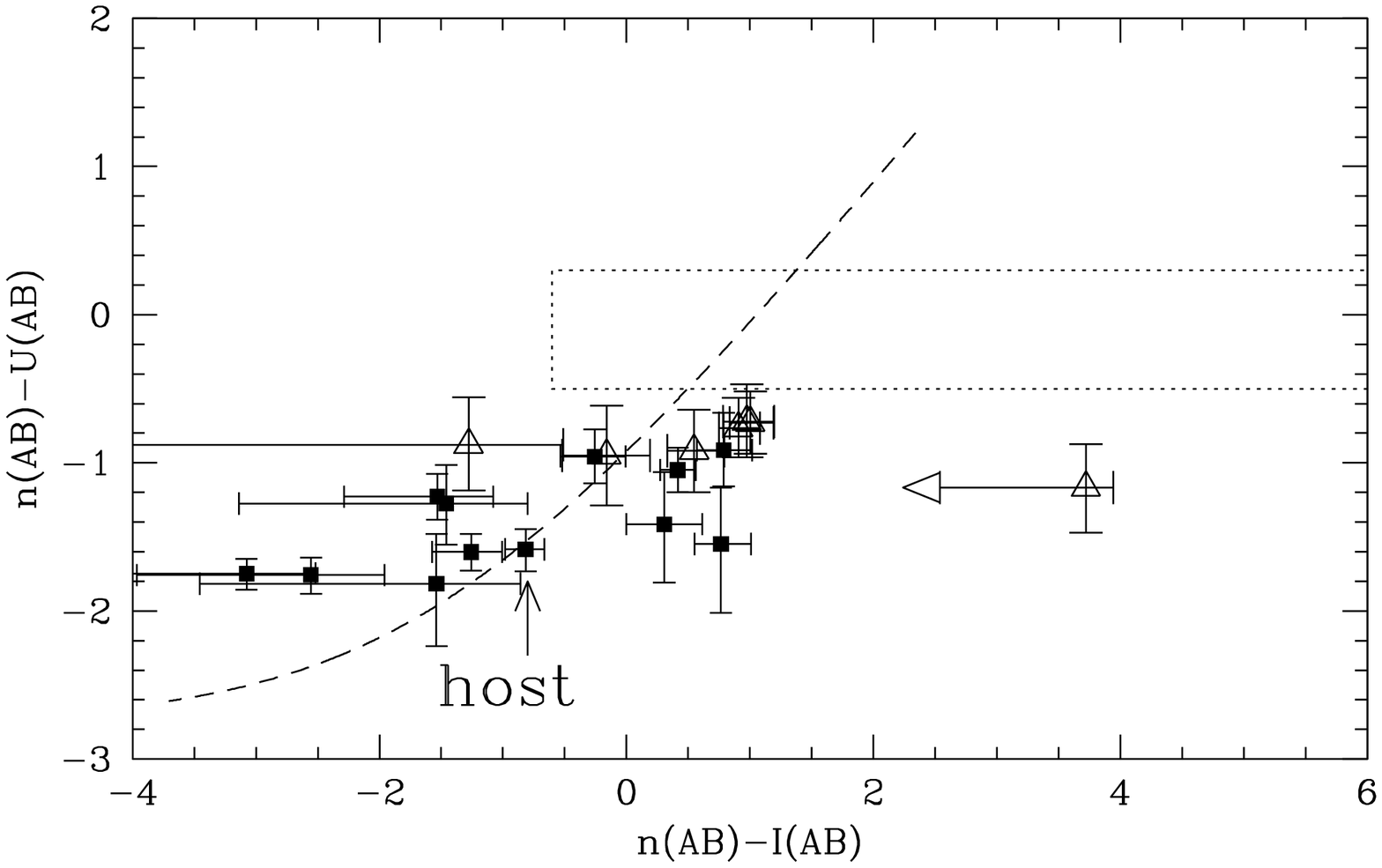, width=9cm, clip=}
\caption{Calculated and observed colour-colour diagrams. 
{\it Upper panel:} 
Calculated colour-colour diagram based on Bruzual and Charlot galaxy
SEDs. The filled squares are 0$<$z$<$1.5 galaxies with ages from a
few to 15 Gyr and the open triangles are 1.5$<$z$<$3.0 galaxies
with ages from a few Myr to 1Gyr. The dotted box contains all these
calculated galaxy colours. The dashed line is the locus of objects
having the same broad-band colours as GRB~000926 host galaxy and various 
amounts of absorption (upper part) or emission (lower part) in the 
narrow-band filter.
{\it Middle panel:} 
Colour-colour diagrams for all objects in the two GRB fields. The 
squares with error-bars indicate objects detected at S/N$>$5 in the 
narrow-band image. As expected, most objects have colours consistent
with being in the dotted box, however, a number of objects, including 
the GRB~000926 host, are seen in the lower left part of the 
diagram.
{\it Lower panel:} 
The colours of the 19 candidate LEGOs. The rank 1 and 2 candidates are
shown with filled squares and open triangles respectively.
}
\label{colcol}
\end{center}
\end{figure}

\subsection{Selection of Candidate LEGOs}
\label{lego}

For the final selection of candidate LEGOs we use the ``narrow minus
on-band-broad'' versus ``narrow minus off-band-broad'' colour/colour
plot technique (M{\o}ller \& Warren 1993; Fynbo et al. 1999, 2000b). 
In order to constrain where objects with no special spectral features in the 
narrow filter fall in the diagram, we calculate colours based on synthetic 
galaxy SEDs taken from the Bruzual \& Charlot 1995 models. 
We have used models with ages ranging from a few Myr to 15 Gyr and with 
redshifts from 0 to 1.5 (open squares) and models with ages ranging from
a few Myr to 1 Gyr with redshifts from 1.5 to 3.0 (open triangles). Higher
redshifts are not relevant for our 3700 \AA \ selected survey. For
the colours of high redshift galaxies we include the effect of
intergalactic Lyman line absorption (M{\o}ller \& Jakobsen 1990; Madau 
1995). In Fig.~\ref{colcol} we show the n(AB)$-$U(AB) 
versus n(AB)$-$I(AB) colour diagram for the calculated  galaxy colours 
(upper panel) and for the observed sources in the two GRB fields
(middle and lower panels). The open squares 
in the upper panel shows the colours of the model SEDs. 
The dashed line indicates where objects with broad-band colour 
equal to that of the host galaxy of GRB~000926, and with either absorption or 
emission in the narrow filter, will fall.
Emission line objects will fall in the lower left corner
of the diagram (due to excess emission in the narrow-filter). 
In the middle panel we show the colour-colour diagram for all objects
detected in the two fields. Several objects are found
to lie significantly away from the locus of continuum objects. 

The probability that a Ly$\alpha$ emitter will separate itself 
significantly from
the pure continuum objects is a (complex) function of its continuum
magnitude, colour, and Ly$\alpha$ emission line equivalent width. It is
therefore necessary to define a strategy to most effciently identify the
Ly$\alpha$ emitters. One may decide to err on the low probability side
(providing a large list of candidates but with fairly low confirmation 
fraction, e.g. $\sim$30\% in the work by Arnaboldi et al. 2001 and
references therein) or on the high probability side (shorter candidate
list, but higher confirmation efficiency, e.g. $\sim$100\% in 
Warren \& M{\o}ller (1996),
Kudritzki et al. (2000), and Fynbo et al. (2001b)). Both strategies
have their merits and their drawbacks; the conservative candidate list
will likely miss a number of objects while the non-conservative candidate
list in the end may provide a few more confirmed LEGOs, but at the price
of a much lower efficiency during spectroscopic confirmation. However,
when spectroscopic follow-up is done via MOS (Multi-Object Spectroscopy) 
one can to some extent have the best of both strategies as the less secure 
candidates
can be used for slits not covered by the secure candidates. We
therefore decided to make a ranked candidate list in the following way:
We first select all objects with S/N$>$5 in the narrow-band image and an 
isophotal colour n(AB)$-$U(AB)$<$$-0.7$. We 
find 32 such objects. We subsequently rank these candidates in four
groups, 1:{\it certain}, 2:{\it  good}, 3:{\it possible}, and
4:{\it rejected}. For the rank 1,2, and 3 groups we expect rough 
confirmation efficiencies of $\sim$100\%, $\sim$75\% and $\sim$50\% 
respectively. We base the rank on visual inspection (to exclude
spurious detections or objects with wrong colours due to bright
neighbours) and on the aperture photometry (the aperture
colour should also be consistent with excess emission in the narrow
filter). In the following we will only discuss the 12 and 7 candidates 
that fall in the {\it certain} and {\it good} categories (our conservative
candidate list). The {\it possible}
candidates we plan to use as fill-up objects during the follow-up
spectroscopic observations. The colours of the rank 1 and 2
candidates are shown in the lower panel of Fig.~\ref{colcol}. 
We find eleven rank 1 and 2 candidate LEGOs in the field of GRB~000301C 
and eight in the field of GRB~000926 (including the host galaxy). 
In Table~\ref{candprop1} and Table~\ref{candprop2} we list 
the photometric properties of the rank 1 and 2 candidates. In
Fig.~\ref{Sgals} we show regions of size 10$\times$10 arcsec$^2$
around each candidate from the combined narrow-band (top row), U-band
(centre row), and I-band (bottom row) images. In Fig.~\ref{fields}
we show the distribution of the candidate LEGOs in each of the two
fields projected on the sky. In the field of GRB~000301C (the left plot)
the candidates are all located in the southern part of the field.
In the field of GRB~000926 the eight galaxies are more uniformly 
distributed over the field of view. The skewed distribution of LEGOs
in the field of GRB~000301C suggests a large scale structure, possibly
a filament similar to the one found in the field of the z=3.04 QSO
Q1205-30 (M\o ller \& Fynbo 2001). Spectroscopic follow-up is needed
to place the LEGOs in 3D redshift space.

Below we describe how we determine Ly$\alpha$ fluxes, Ly$\alpha$
equivalent widths 
and star formation rates for the candidate LEGOs.

\noindent
{\it Ly$\alpha$ fluxes:}
We determine Ly$\alpha$ fluxes, $f_{\rm{Ly\alpha}}$, as 
\[
f_{\rm{Ly\alpha}} = \bar{f_{\lambda}}\cdot
\int{T_{\rm{\rm{CS3701/45}}}(\lambda)d\lambda}/T_{\rm{\rm{CS3701/45}}}^{\rm{max}},
\]
\noindent
where $T_{\rm{\rm{CS3701/45}}}(\lambda)$ is the transmission curve of the narrow 
filter, $T_{\rm{CS3701/45}}^{\rm{max}}$ is the top transmission of the filter, 
the term $\int{T_{\rm{CS3701/45}}(\lambda)d\lambda}/T_{\rm{CS3701/45}}^{\rm{max}}$
is the ``effective'' width of the filter (equal to 45 \AA), and
$\bar{f_{\lambda}}$ is the average specific flux of the source
determined from the narrow-band AB magnitude, n(AB), as 
\[
\bar{f_{\lambda}} = \bar{f_{\nu}}\frac{c}{\lambda^2}
= 10^{-0.4(n({\rm AB})+48.60)}\frac{c}{\lambda^2}
\]

\noindent
{\it Ly$\alpha$ equivalent widths:} To estimate equivalent widths we 
derive the relation between the n(AB)-U(AB) colour and
the observed equivalent width (W$_{\rm{em}}(obs)$). The n(AB)-U(AB) colour
for an object with specific flux $f_{\nu}$ is given by the expression:
\[
-2.5\log{\frac
{\int{f_{\nu}(\nu)T_{\rm{CS3701/45}}(\nu)d{\nu}}/\int{T_{\rm{CS3701/45}}(\nu)d{\nu}}}
{\int{f_{\nu}(\nu)T_{\rm{U}}(\nu)d{\nu}}/\int{T_{\rm{U}}(\nu)d{\nu}}},
}
\]
where $T_{\rm{CS3701/45}}$ and $T_{\rm{U}}$ are the transmission curves of the 
narrow and broad-band U filters respectively. We assume that the 
specific flux $f_{\lambda}$ is a constant, $f_0$,   
except for an emission line with Gaussian profile at 
$\lambda_0 = 3701$ \AA \ with equivalent width $W_{\rm{em}}(obs)$. 
Under these assumptions we have
\[
f_{\lambda} = f_0 + f_0 G(\lambda-\lambda_0) \times W_{\rm{em}}(\rm{obs}),
\]
where $G(\lambda-\lambda_0)$ is a Gaussian function with area equal 1.

Finally, we use $f_{\nu} \propto f_{\lambda} \lambda^2$ and insert in the
expression for n(AB)-U(AB) above to obtain the relation between W$_{em}$(obs)
and n(AB)-U(AB). Assuming instead that $f_{\nu}$ is constant leads to
practically the same relation.

\noindent
{\it Star formation rates:}
Kennicutt (1983) has derived the following scaling relation between 
the Star Formation Rate (SFR) and the H$\alpha$ luminosity L(H$\alpha$):
\[
\rm{SFR} = L(H\alpha)/1.12\times10^{41} erg s^{-1}
\]
Assuming negligible dust extinction we can use the case B recombination 
(relevant for gas that is optically thick in the \ion{H}{i} resonance 
lines) ratio L(Ly$\alpha$)/L(H$\alpha$)=8.7 (Brocklehurst 1971) to derive 
L(H$\alpha$), where 
L(Ly$\alpha$) = $f_{\rm{Ly\alpha}}4\pi d_{\rm{lum}}^2$, and hence the SFR.
The effects of dust will be to reduce the Ly$\alpha$ flux for a given
SFR (Charlot \& Fall 1993; Valls-Gabaud 1993; Deharveng et al. 1995; see 
also Neufeld 1991). Therefore SFRs derived in this way must be considered 
lower limits.

The Ly$\alpha$ fluxes, Ly$\alpha$ equivalent widths and SFRs
for the candidate LEGOs appear in Table~\ref{candprop1} and
Table~\ref{candprop2}.

\subsection{Extendedness of Ly$\alpha$ emission}
\label{extended}

\begin{figure}
\begin{flushleft}
\epsfig{file=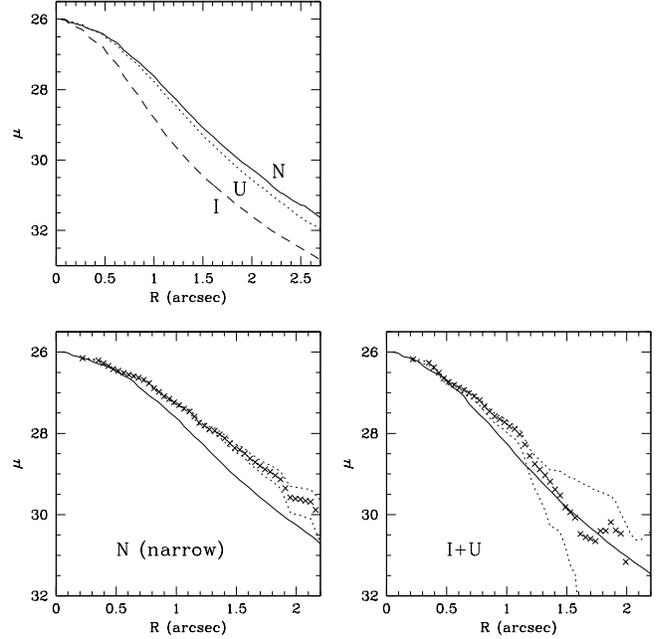, width=8.7cm}
\caption{
{\it Top left:} PSF (calculated as azimuthal average) of the combined
images of I, U and narrow. 
{\it Bottom left:} Narrow-band image azimuthal average of stacked point
sources (solid line) and of stacked Ly$\alpha$ emitters (LEGOs, shown
with $\times$). The dotted lines indicate the $\pm 1\sigma$ range. The 
LEGOs are clearly extended. 
{\it Bottom right:} Broad-band  (sum of I and U) image azimuthal average
of stacked point sources (solid line) and of stacked LEGOs ($\times$). 
The dotted lines indicate the $\pm 1\sigma$ range. The S/N of the broad 
band profile 
is lower than for the narrow-band profile, but there is marginal
evidence that the continuum emission of the LEGOs are less extended 
than the Ly$\alpha$ emission. 
}
\label{profile}
\end{flushleft}
\end{figure}

\begin{figure*}
\begin{center}
\epsfig{file=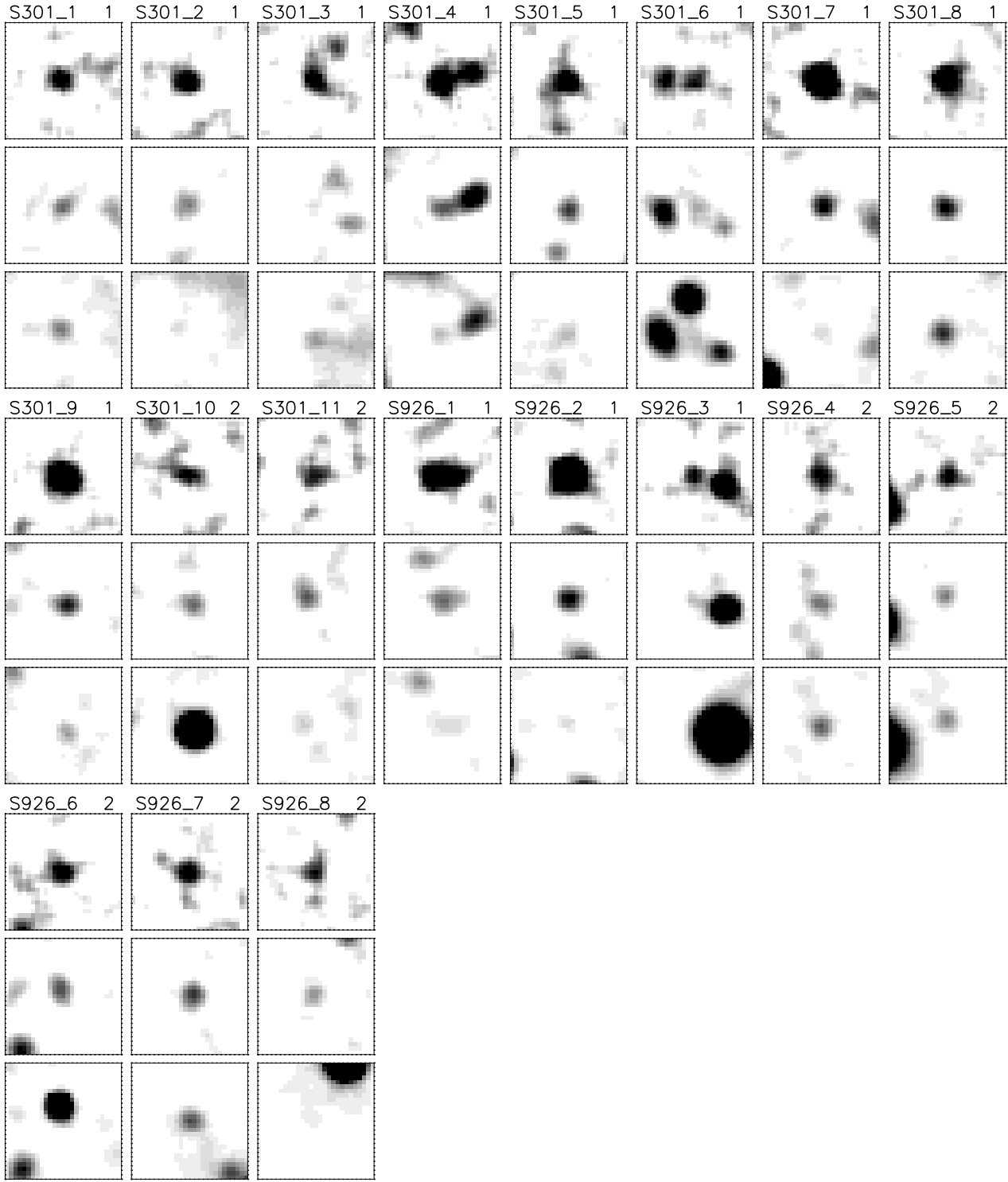, width=16.7cm}
\caption{Regions of size 10$\times$10 arcsec$^2$ around each of the
eleven candidates LEGOs in the field of GRB~000301C and the eight 
candidates LEGOs in the field of GRB~000926. For each candidate
we show a sub-image from all three filters narrow (top row), U (middle
row) and I (bottom row). The name and rank of 
the candidates are printed above each of the narrow-band sub-images.
The host galaxy of GRB~000926 is S926\_1.}
\label{Sgals}
\end{center}
\end{figure*}

Previous studies have found evidence that the Ly$\alpha$ emission
is more extended than the continuum emission (M\o ller \& Warren 1998;
Fynbo et al. 2001). To test whether this is also the case for the
Ly$\alpha$ emitters in this study we proceed as follows. Because the
individual sources are too faint for precise profile fits, we stack the
isolated rank 1 candidates and measure the radial profile on the
stacked image in narrow-band, U, and I.
We compare the profiles with the profile of the same
number of point sources stacked in the same way. The results of this
procedure are shown in the three panels of Fig.~\ref{profile}. The
Ly$\alpha$ objects are clearly extended. Furthermore, although the S/N
is lower than for the narrow-band profile, there is marginal evidence
that the continuum sources of the Ly$\alpha$ objects are
less extended than the Ly$\alpha$ emission itself.

We finally use the profiles of the point-sources and the candidate 
Ly$\alpha$ emitters to determine the aperture corrections for our
3\farcs0 and 4\farcs0 circular apertures relative to a 11 circular
aperture (as used for the standard stars, Sect.~\ref{obs}). We find 
aperture corrections 
for the 3\farcs0(4\farcs0) apertures of 0.36(0.21), 0.27(0.15) and 
0.23(0.14) mag for narrow, U and I respectively.

\subsection{Narrow-band absorption objects}
\label{abs}
Roughly half of the Lyman-Break selected Galaxies have Ly$\alpha$ in
{\it absorption} (e.g. Shapley et al. 2001, their Fig.~7). 
In order to detect z=2.04 galaxies with absorption in the narrow-band 
filter, we run SExtractor using the combined U-band image as
detection image, hence optimising the detection of objects that are
bright in the U-band. We detect 515 and 578 objects above S/N=5
in isophotal apertures in the fields of GRB~000301C and GRB~000926
respectively. Since the observations were tuned to select objects
with excess emission in the narrow-band filter the combined narrow-band 
image is about 1 magnitude shallower than the combined U-band image
for objects with flat SEDs. We can therefore only detect a {\it deficit} 
of emission in the narrow filter for relatively bright U-band selected 
objects (brighter than U(AB)=25.3). As candidates we select galaxies with 
colours above n(AB)$-$U(AB)=0.6 within 1$\sigma$. We detect two and five 
such galaxies in the field of GRB~000301C and GRB~000926 respectively. 
Their positions are indicated with squares in Fig.~\ref{fields}. Deep
follow-up spectroscopy is required to confirm if these objects are high
redshift galaxies with strong z=2.04 intrinsic or intervening
Ly$\alpha$ absorption lines.

%=====================Begin Table 2==============================
\begin{table*}[ht]
\begin{center}
\caption{Photometric properties of the 12 candidate LEGOs of rank
1 in the
fields of GRB~000301C and GRB~000926. Aperture corrections for
the 3\farcs0(4\farcs0) aperture are 0.36(0.21), 0.27(0.15) and 
0.23(0.14) mag for narrow, U and I respectively.}
\begin{tabular}{@{}lccccccc}
\hline
Object  & n(AB)  & U(AB)  &  I(AB)  &  f(Ly$\alpha$) & W$_{\rm{em}}$(obs) & SFR$_{\rm{Ly\alpha}}$  \\
         & &        &       &  10$^{-17}$ erg s$^{-1}$ cm$^{-2}$ & {\AA} &
M$_{\sun}$ yr$^{-1}$ \\
\hline
{\it Isophote} \\
S301\_1 & 25.07$^{+0.21}_{-0.17}$ & 25.98$^{+0.17}_{-0.15}$ &
24.28$^{+0.13}_{-0.11}$ &   3.36$\pm$  0.54 &  69$^{+  38}_{-  28}$ &
0.92$\pm$0.15 \\
S301\_2 & 24.88$^{+0.18}_{-0.16}$ & 26.16$^{+0.23}_{-0.19}$ &
26.34$^{+1.66}_{-0.63}$ &   3.98$\pm$  0.58 &  130$^{+  75}_{-  47}$ &
1.09$\pm$0.16 \\
S301\_3 & 24.98$^{+0.18}_{-0.15}$ & 26.80$^{+0.39}_{-0.29}$ &
26.52$^{+1.89}_{-0.65}$ &   3.64$\pm$  0.52 &  325$^{+ 464}_{- 144}$ &
1.00$\pm$0.14 \\
S301\_4$^a$ & 23.91$^{+0.11}_{-0.10}$ & 25.13$^{+0.12}_{-0.11}$ &
25.44$^{+0.75}_{-0.44}$ &   9.80$\pm$  0.87 &  120$^{+  36}_{-  27}$ &
2.68$\pm$0.24 \\
S301\_5 & 24.54$^{+0.14}_{-0.13}$ & 25.50$^{+0.13}_{-0.12}$ &
24.80$^{+0.25}_{-0.20}$ &   5.46$\pm$  0.64 &   75$^{+  28}_{-  22}$ &
1.49$\pm$0.18 \\
S301\_6$^a$ & 25.48$^{+0.24}_{-0.20}$ & 26.90$^{+0.34}_{-0.26}$ &
25.17$^{+0.24}_{-0.19}$ &   2.30$\pm$  0.42 &  163$^{+ 155}_{-  73}$ &
0.63$\pm$0.11 \\
S301\_7 & 23.11$^{+0.06}_{-0.06}$ & 24.87$^{+0.11}_{-0.10}$ &
25.67$^{+1.41}_{-0.59}$ &  20.40$\pm$  1.05 &  292$^{+  75}_{-  54}$ &
5.58$\pm$0.29 \\
S301\_8 & 24.13$^{+0.11}_{-0.10}$ & 25.18$^{+0.11}_{-0.10}$ &
23.72$^{+0.10}_{-0.09}$ &   7.95$\pm$  0.75 &   88$^{+  26}_{-  21}$ &
2.18$\pm$0.21 \\
S301\_9 & 23.30$^{+0.07}_{-0.06}$ & 24.90$^{+0.11}_{-0.10}$ &
24.56$^{+0.31}_{-0.24}$ &  17.14$\pm$  1.00 &  223$^{+  55}_{-  41}$ &
4.69$\pm$0.27 \\
{\it 3\farcs0 aperture} \\
S301\_1 & 24.58$^{+0.26}_{-0.21}$ & 25.49$^{+0.22}_{-0.18}$ &
23.88$^{+0.17}_{-0.15}$ &   5.27$\pm$  1.04 & - & 1.44$\pm$0.28 \\
S301\_2 & 24.29$^{+0.20}_{-0.17}$ & 25.27$^{+0.18}_{-0.15}$ &
25.92$^{+3.90}_{-0.74}$ &   6.89$\pm$  1.06 & - & 1.88$\pm$0.29 \\
S301\_3 & 24.32$^{+0.20}_{-0.17}$ & 26.15$^{+0.45}_{-0.32}$ &
25.92$^{+3.90}_{-0.74}$ &   6.72$\pm$  1.06 & - & 1.84$\pm$0.29 \\
S301\_4$^a$ & 23.74$^{+0.11}_{-0.10}$ & 25.07$^{+0.15}_{-0.13}$ &
25.82$^{+2.36}_{-0.69}$ &  11.39$\pm$  1.09 & - & 3.12$\pm$0.30 \\
S301\_5 & 24.16$^{+0.17}_{-0.15}$ & 25.10$^{+0.15}_{-0.13}$ &
24.30$^{+0.27}_{-0.21}$ &   7.73$\pm$  1.07 & - & 2.12$\pm$0.29 \\
S301\_6$^a$ & 24.43$^{+0.23}_{-0.19}$ & 25.52$^{+0.23}_{-0.19}$ &
23.94$^{+0.19}_{-0.16}$ &   6.05$\pm$  1.05 & - & 1.66$\pm$0.29 \\
S301\_7 & 23.11$^{+0.06}_{-0.06}$ & 24.83$^{+0.12}_{-0.10}$ &
25.92$^{+3.90}_{-0.74}$ &  20.47$\pm$  1.11 & - & 5.60$\pm$0.30 \\
S301\_8 & 23.87$^{+0.13}_{-0.12}$ & 24.86$^{+0.12}_{-0.11}$ &
23.39$^{+0.11}_{-0.10}$ &  10.14$\pm$  1.08 & - & 2.77$\pm$0.30 \\
S301\_9 & 23.27$^{+0.07}_{-0.07}$ & 24.90$^{+0.12}_{-0.11}$ &
24.64$^{+0.39}_{-0.28}$ &  17.54$\pm$  1.10 & - & 4.80$\pm$0.30 \\
{\it 4\farcs0 aperture} \\
S301\_1 & 24.59$^{+0.34}_{-0.26}$ & 25.27$^{+0.24}_{-0.20}$ &
23.91$^{+0.24}_{-0.20}$ &   5.20$\pm$  1.26 & - & 1.42$\pm$0.34 \\
S301\_2 & 24.33$^{+0.26}_{-0.21}$ & 25.09$^{+0.20}_{-0.17}$ &
25.62$^{+4.20}_{-0.74}$ &   6.64$\pm$  1.28 & - & 1.82$\pm$0.35 \\
S301\_3 & 24.14$^{+0.21}_{-0.18}$ & 25.94$^{+0.51}_{-0.34}$ &
25.62$^{+4.20}_{-0.74}$ &   7.87$\pm$  1.30 & - & 2.16$\pm$0.36 \\
S301\_4$^a$ & 23.59$^{+0.12}_{-0.11}$ & 24.73$^{+0.14}_{-0.12}$ &
25.62$^{+4.20}_{-0.74}$ &  13.10$\pm$  1.33 & - & 3.59$\pm$0.37 \\
S301\_5 & 24.08$^{+0.20}_{-0.17}$ & 25.14$^{+0.21}_{-0.18}$ &
24.01$^{+0.27}_{-0.22}$ &   8.33$\pm$  1.30 & - & 2.28$\pm$0.36 \\
S301\_6$^a$ & 24.24$^{+0.23}_{-0.19}$ & 25.10$^{+0.20}_{-0.17}$ &
23.10$^{+0.11}_{-0.10}$ &   7.23$\pm$  1.29 & - & 1.98$\pm$0.35 \\
S301\_7 & 23.02$^{+0.07}_{-0.07}$ & 24.77$^{+0.15}_{-0.13}$ &
25.63$^{+4.19}_{-0.74}$ &  22.09$\pm$  1.36 & - & 6.05$\pm$0.37 \\
S301\_8 & 23.75$^{+0.14}_{-0.13}$ & 24.77$^{+0.15}_{-0.13}$ &
23.42$^{+0.15}_{-0.13}$ &  11.26$\pm$  1.33 & - & 3.08$\pm$0.36 \\
S301\_9 & 23.15$^{+0.08}_{-0.07}$ & 24.84$^{+0.16}_{-0.14}$ &
24.84$^{+0.70}_{-0.42}$ &  19.75$\pm$  1.36 & - & 5.41$\pm$0.37 \\
{\it Isophote} \\
S926\_1$^b$ & 23.35$^{+0.07}_{-0.07}$ & 24.94$^{+0.13}_{-0.12}$ &
24.17$^{+0.15}_{-0.13}$ &  16.30$\pm$  1.02 &  217$^{+  62}_{-  45}$ &
4.46$\pm$0.28 \\
S926\_2 & 22.78$^{+0.05}_{-0.04}$ & 24.52$^{+0.10}_{-0.09}$ &
25.85$^{+1.21}_{-0.56}$ &  27.73$\pm$  1.14 &  287$^{+  62}_{-  46}$ &
7.59$\pm$0.31 \\
S926\_3$^a$ & 25.50$^{+0.23}_{-0.19}$ & 27.04$^{+0.42}_{-0.30}$ &
24.73$^{+0.10}_{-0.09}$ &   2.27$\pm$  0.40 &  204$^{+ 268}_{-  96}$ &
0.62$\pm$0.11 \\
{\it 3\farcs0 aperture} \\
S926\_1$^b$ & 23.48$^{+0.08}_{-0.08}$ & 24.94$^{+0.13}_{-0.12}$ &
24.24$^{+0.17}_{-0.14}$ &  14.49$\pm$  1.04 & - & 3.97$\pm$0.28 \\
S926\_2 & 22.83$^{+0.05}_{-0.04}$ & 24.60$^{+0.10}_{-0.09}$ &
25.57$^{+0.72}_{-0.43}$ &  26.39$\pm$  1.06 & - & 7.22$\pm$0.29 \\
S926\_3$^a$ & 24.91$^{+0.35}_{-0.27}$ & 25.66$^{+0.28}_{-0.22}$ &
22.16$^{+0.02}_{-0.02}$ &   3.88$\pm$  0.97 & - & 1.06$\pm$0.26 \\
{\it 4\farcs0 aperture} \\
S926\_1$^b$ & 23.23$^{+0.09}_{-0.08}$ & 24.75$^{+0.15}_{-0.13}$ &
23.81$^{+0.15}_{-0.13}$ &  18.26$\pm$  1.38 & - & 5.00$\pm$0.38 \\
S926\_2 & 22.70$^{+0.05}_{-0.05}$ & 24.57$^{+0.13}_{-0.11}$ &
25.89$^{+2.10}_{-0.67}$ &  29.67$\pm$  1.40 & - & 8.12$\pm$0.38 \\
S926\_3$^a$ & 24.68$^{+0.38}_{-0.28}$ & 25.60$^{+0.36}_{-0.27}$ &
21.60$^{+0.02}_{-0.02}$ &   4.80$\pm$  1.27 & - & 1.31$\pm$0.35 \\
\hline
\label{candprop1}
\end{tabular}
\end{center}
\begin{footnotesize}
\vskip -0.6cm
\noindent 
$^a$ Broad-band photometry affected by bright nearby object.

\noindent
$^b$ The host galaxy of GRB~000926. 
\end{footnotesize}
\end{table*}
%=====================End Table 2===============================

%=====================Begin Table 2==============================
\begin{table*}[ht]
\begin{center}
\caption{Photometric properties of the 7 candidate LEGOs of rank 2
in the
fields of GRB~000301C and GRB~000926. Aperture corrections for
the 3\farcs0(4\farcs0) aperture are 0.36(0.21), 0.27(0.15) and 
0.23(0.14) mag for narrow, U and I respectively.}
\begin{tabular}{@{}lccccccc}
\hline
Object  & n(AB)  & U(AB)  &  I(AB)  &  f(Ly$\alpha$) & W$_{\rm{em}}$(obs) & SFR$_{\rm{Ly\alpha}}$ \\
        &   &        &       &  10$^{-17}$ erg s$^{-1}$ cm$^{-2}$ & {\AA} &
M$_{\sun}$ yr$^{-1}$ \\
\hline
{\it Isophote} \\
S301\_10$^a$ & 25.40$^{+0.22}_{-0.18}$ & 26.57$^{+0.24}_{-0.20}$ &
21.67$^{+0.01}_{-0.01}$ &   2.48$\pm$  0.42 &  108$^{+  71}_{-  44}$ &
0.68$\pm$0.11 \\
S301\_11 & 25.74$^{+0.27}_{-0.21}$ & 26.62$^{+0.22}_{-0.18}$ &
27.02$^{+2.80}_{-0.71}$ &   1.81$\pm$  0.36 &   65$^{+  47}_{-  32}$ &
0.49$\pm$0.10 \\
{\it 3\farcs0 aperture} \\
S301\_10$^a$ & 24.67$^{+0.29}_{-0.23}$ & 25.58$^{+0.24}_{-0.20}$ &
20.54$^{+0.01}_{-0.01}$ &   4.83$\pm$  1.03 & - & 1.32$\pm$0.28 \\
S301\_11 & 24.55$^{+0.26}_{-0.21}$ & 25.49$^{+0.22}_{-0.18}$ &
25.07$^{+0.64}_{-0.40}$ &   5.41$\pm$  1.04 & - & 1.48$\pm$0.29 \\
{\it 4\farcs0 aperture} \\
S301\_10$^a$ & 24.68$^{+0.38}_{-0.28}$ & 25.74$^{+0.40}_{-0.29}$ &
20.43$^{+0.01}_{-0.01}$ &   4.79$\pm$  1.25 & - & 1.31$\pm$0.34 \\
S301\_11 & 24.49$^{+0.30}_{-0.24}$ & 25.24$^{+0.23}_{-0.19}$ &
24.79$^{+0.66}_{-0.41}$ &   5.72$\pm$  1.26 & - & 1.57$\pm$0.35 \\
{\it Isophote} \\
S926\_4 & 25.04$^{+0.21}_{-0.18}$ & 25.76$^{+0.16}_{-0.14}$ &
24.06$^{+0.08}_{-0.07}$ &   3.46$\pm$  0.56 &   47$^{+  28}_{-  22}$ &
0.95$\pm$0.15 \\
S926\_5 & 25.33$^{+0.22}_{-0.18}$ & 26.25$^{+0.21}_{-0.17}$ &
24.78$^{+0.12}_{-0.11}$ &   2.64$\pm$  0.44 &   70$^{+  44}_{-  30}$ &
0.72$\pm$0.12 \\
S926\_6 & 24.72$^{+0.18}_{-0.15}$ & 25.45$^{+0.14}_{-0.12}$ &
23.71$^{+0.06}_{-0.06}$ &   4.64$\pm$  0.65 &   48$^{+  24}_{-  19}$ &
1.27$\pm$0.18 \\
S926\_7 & 24.79$^{+0.16}_{-0.14}$ & 25.56$^{+0.13}_{-0.12}$ &
23.89$^{+0.06}_{-0.06}$ &   4.34$\pm$  0.57 &   52$^{+  24}_{-  19}$ &
1.19$\pm$0.16 \\
S926\_8 & 25.64$^{+0.27}_{-0.21}$ & 26.59$^{+0.26}_{-0.21}$ &
25.80$^{+0.29}_{-0.23}$ &   1.98$\pm$  0.39 &   74$^{+  58}_{-  37}$ &
0.54$\pm$0.11 \\
{\it 3\farcs0 aperture} \\
S926\_4 & 24.46$^{+0.22}_{-0.18}$ & 25.11$^{+0.16}_{-0.14}$ &
23.74$^{+0.10}_{-0.09}$ &   5.91$\pm$  1.00 & - & 1.62$\pm$0.27 \\
S926\_5 & 24.81$^{+0.32}_{-0.24}$ & 25.65$^{+0.28}_{-0.22}$ &
23.94$^{+0.12}_{-0.11}$ &   4.28$\pm$  0.97 & - & 1.17$\pm$0.27 \\
S926\_6 & 24.27$^{+0.18}_{-0.16}$ & 25.14$^{+0.16}_{-0.14}$ &
22.40$^{+0.03}_{-0.03}$ &   7.00$\pm$  1.01 & - & 1.91$\pm$0.28 \\
S926\_7 & 24.48$^{+0.23}_{-0.19}$ & 24.91$^{+0.13}_{-0.12}$ &
23.48$^{+0.08}_{-0.07}$ &   5.76$\pm$  0.99 & - & 1.58$\pm$0.27 \\
S926\_8 & 24.75$^{+0.30}_{-0.23}$ & 25.61$^{+0.27}_{-0.21}$ &
24.95$^{+0.35}_{-0.26}$ &   4.50$\pm$  0.98 & - & 1.23$\pm$0.27 \\
{\it 4\farcs0 aperture} \\
S926\_4 & 24.39$^{+0.28}_{-0.22}$ & 24.97$^{+0.19}_{-0.16}$ &
23.62$^{+0.12}_{-0.11}$ &   6.25$\pm$  1.30 & - & 1.71$\pm$0.36 \\
S926\_5 & 24.83$^{+0.45}_{-0.32}$ & 25.59$^{+0.36}_{-0.27}$ &
23.95$^{+0.17}_{-0.15}$ &   4.19$\pm$  1.26 & - & 1.15$\pm$0.34 \\
S926\_6 & 24.21$^{+0.23}_{-0.19}$ & 24.91$^{+0.18}_{-0.15}$ &
21.94$^{+0.02}_{-0.02}$ &   7.37$\pm$  1.32 & - & 2.02$\pm$0.36 \\
S926\_7 & 24.48$^{+0.31}_{-0.24}$ & 24.84$^{+0.17}_{-0.14}$ &
23.20$^{+0.08}_{-0.08}$ &   5.75$\pm$  1.29 & - & 1.57$\pm$0.35 \\
S926\_8 & 24.60$^{+0.35}_{-0.26}$ & 25.88$^{+0.50}_{-0.34}$ &
24.53$^{+0.30}_{-0.24}$ &   5.20$\pm$  1.28 & - & 1.42$\pm$0.35 \\
\hline
\label{candprop2}
\end{tabular}
\end{center}
\begin{footnotesize}
\vskip -0.6cm
\noindent 
$^a$ Broad-band photometry affected by bright nearby object.
\end{footnotesize}
\end{table*}
%=====================End Table 2===============================

\subsection{The host galaxies}

\begin{figure}
\begin{center}
\epsfig{file=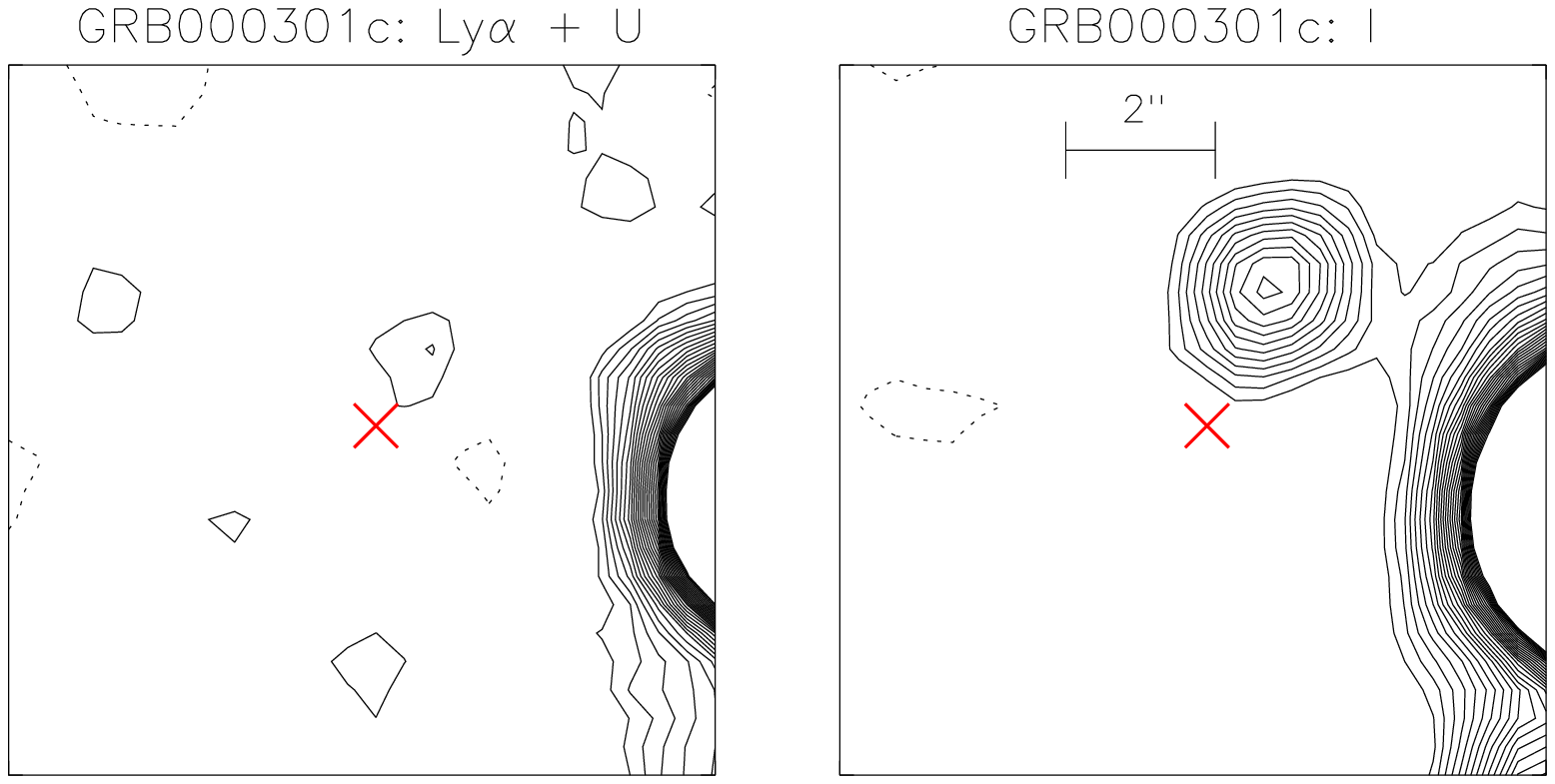, width=9cm}
\epsfig{file=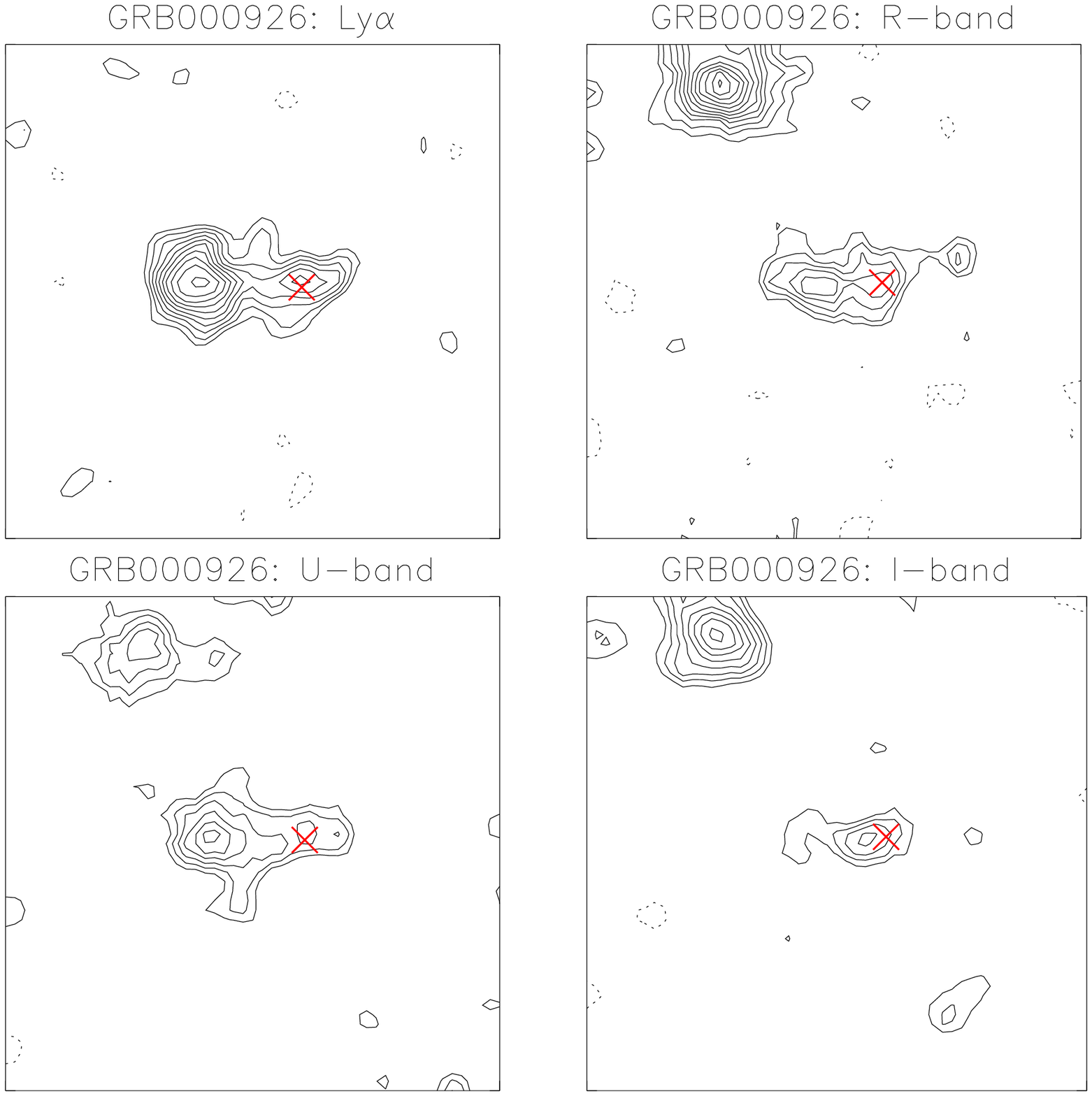, width=9cm}
\caption{
{\it Upper plot:} The position of GRB~000301C as imaged in the I-band
({\it right}) and Ly$\alpha$ + U-band ({\it left}). The size of the
images is 10$\times$10 arcsec$^2$. No significant emission at the 
position of the afterglow is detected to a 2$\sigma$ limit of 
U(AB)=27.7 per arcsec$^2$. 
{\it Lower plot:} The host galaxy of GRB~000926 as imaged in 
Ly$\alpha$, U, R and I. The R-band image is taken from Fynbo et al. 
(2001b). East is to the left and north is up. The size of the images 
is 10$\times$10 arcsec$^2$. The position of the optical afterglow is 
indicated with an $\times$. 
}
\label{host}
\end{center}
\end{figure}

\begin{figure}
\begin{center}
\epsfig{file=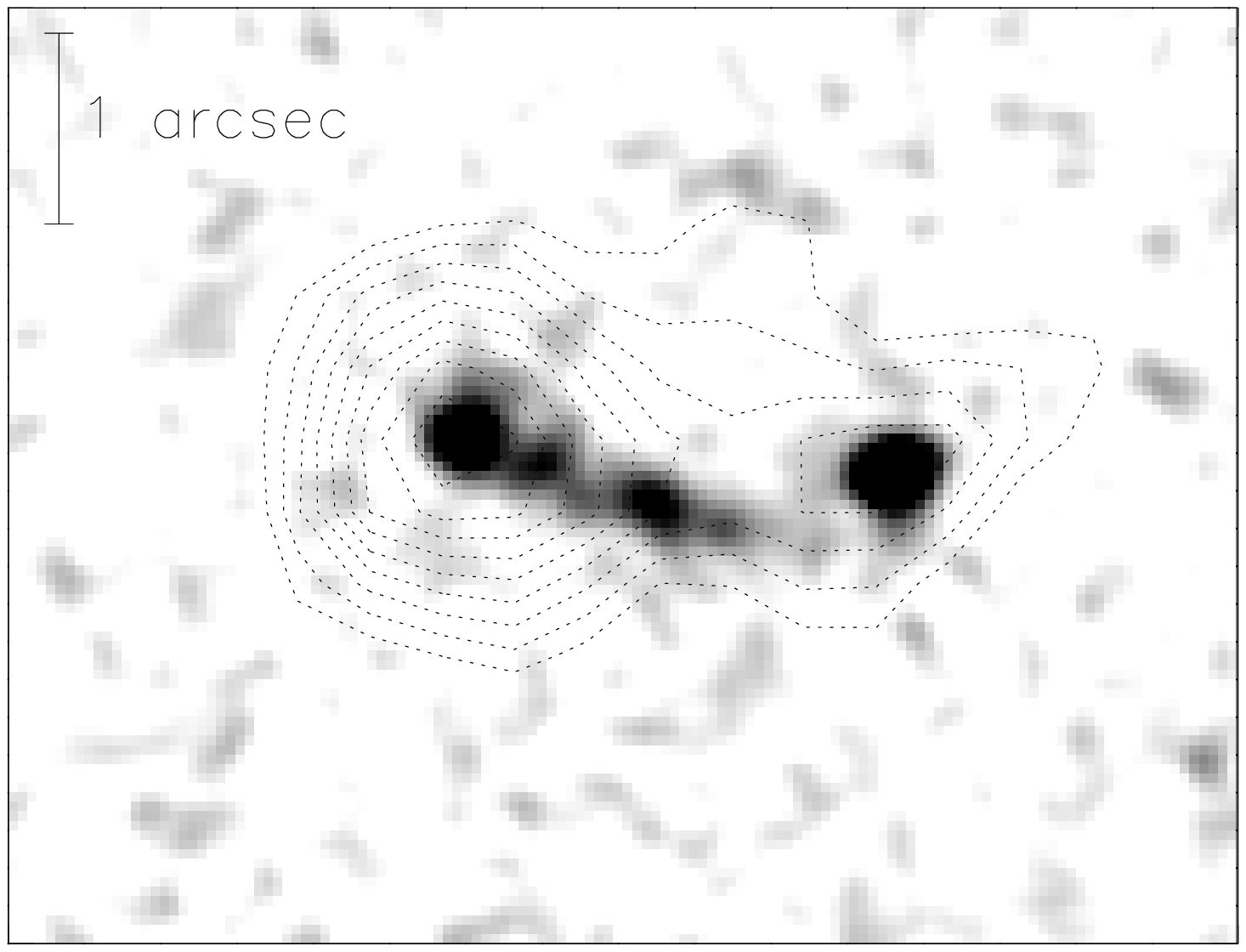, width=8.5cm}
\epsfig{file=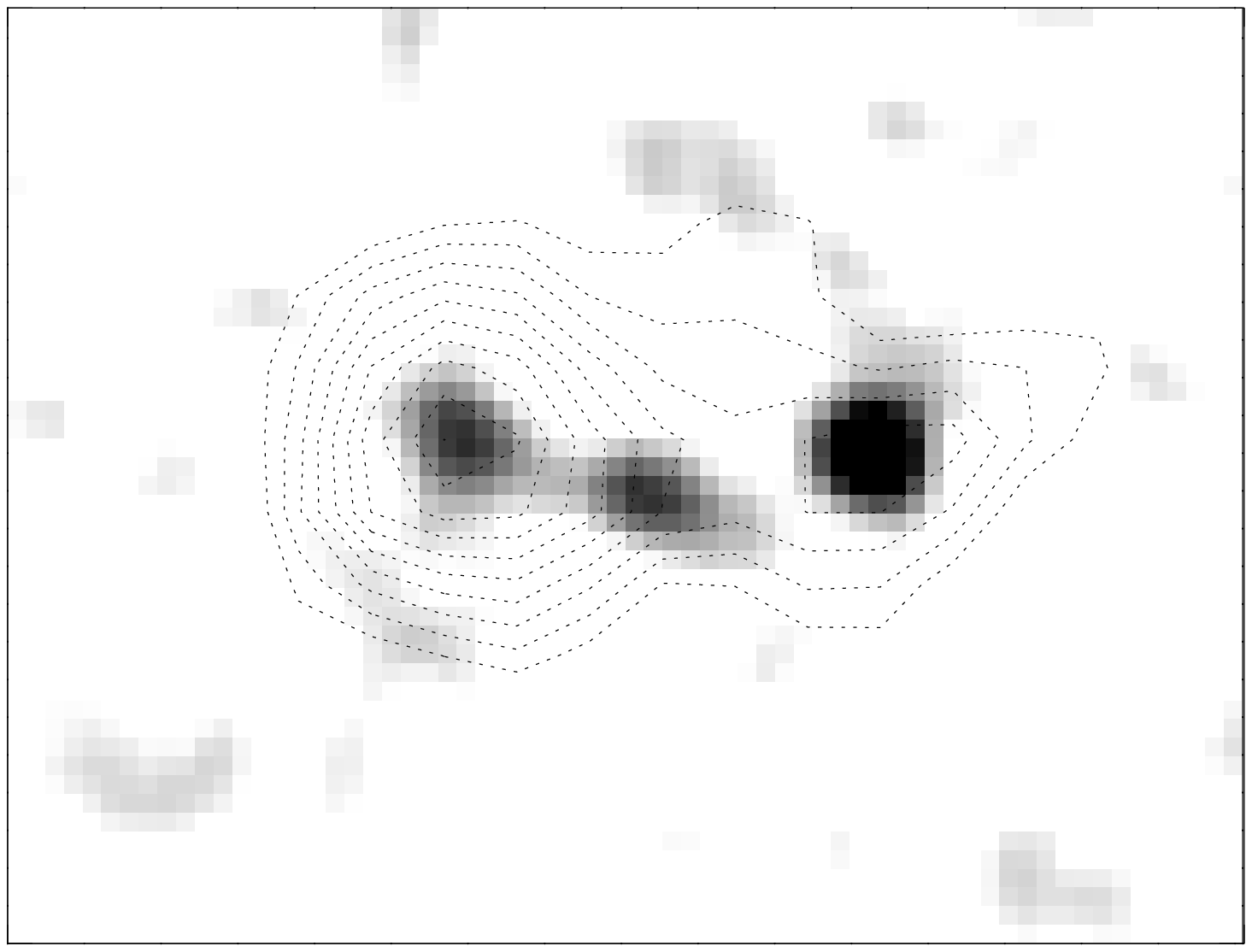, width=8.5cm}
\caption{{\it Top panel:} The WFPC2/F606W drizzled image of the host galaxy 
of GRB~000926 from Castro et al.  (2001) plotted together
with the contours (shown with dotted lines) of the Ly$\alpha$ emission. 
North is up and east is to the left. The scale of the image is indicated 
with a bar in the upper left corner and the estimated error on the alignment 
of the HST and NOT images is shown in the lower left corner. The
Ly$\alpha$ emission covers all the components of the continuum emission
showing that they are all part of the host galaxy and/or its
environment. Most of the Ly$\alpha$ emission (about 65\%) is emitted from 
the eastern most compact knot seen in the F606W image.
{\it Lower panel:} The same region as shown above, but from the (undrizzled) 
F814W image.
}
\label{hsthost}
\end{center}
\end{figure}

We now discuss what can be inferred on the properties of the
host galaxies from the observations presented here.
The host galaxy of GRB~000301C is not detected in any of the bands. 
In Fig.~\ref{host} we show a 10$\times$10 arcsec$^2$ region centred 
on the position of the afterglow of GRB~000301C from the combined I-band
image and from a weighted sum of the combined Ly$\alpha$ and U-band images. 
We calculate the weight as the inverse of the variance of the background
after scaling to the same flux level.
In the I-band we detect the red galaxy 2 arcsec north of the afterglow
also seen in the STIS images by Smette et al. (2001). In the Ly$\alpha$
+ U-band image we detect no significant emission from either this red
galaxy or from the host of GRB~000301C found by Fruchter et al. (2001a)
above an estimated 2$\sigma$ limit of U$\approx$27.7 per arcsec$^2$.
Since the host is very faint (R=28--30, Bloom et al. 2002; 
Levan private communication; Fruchter et al. in prep) this non-detection
is not surprising. 

The host galaxy of GRB~000926 is detected in all bands and is a strong 
Ly$\alpha$ emitter with a rest-frame equivalent width of the 
Ly$\alpha$ emission line of 71 \AA. The galaxy is shown in more detail
in Fig.~\ref{host}. Deep HST images of the host 
has been reported by Price et al. (2001) and Castro et al. (2001). These 
images confirm what was found from the ground based R-band images  
(Fynbo et al. 2001b and Fig.~\ref{host}), namely that there are several
compact objects within a few arcsec, one of which coincide with the position 
of the OT. To better compare the extent of the Ly$\alpha$ emission to
that of the continuum we retrieved and drizzled (Fruchter \& Hook 2002)
the May 19 2001 F606W HST imaging data of the host from the HST
archive. These data are published in Castro et al. (2001). In the top panel 
of Fig.~\ref{hsthost} we show a section of the F606W image
together with the contours of the Ly$\alpha$ emission. The 
Ly$\alpha$ emission covers all of the components seen in the
continuum and we conclude that they are all at the same redshift
and hence part of the host galaxy and/or its environment.
As a simple model we fit the Ly$\alpha$ image with two point sources
using DAOPHOT (Stetson 1987, 1997). The derived distance between the two best
fitting point sources is 2.22$\pm$0.08 arcsec (corresponding to 20.0 kpc), 
which is consistent with the distance
between the brightest knots in the HST image (2.25$\pm$0.01 arcsec). From 
the point source fit we find that about 65\% of the Ly$\alpha$ emission 
is emitted from the eastern compact knot and 35\% from the western knot.
The position of the western (Ly$\alpha$) knot is spatially coincident with the
position of the OT to within 2$\sigma$ of our error of about 0.10 arcsec
on the relative astrometry between the R-band image with the OT from 
Fynbo et al. (2001a) and the Ly$\alpha$ image. Castro et al. (2001) measure 
an impact parameter of the OT
relative to the centroid of the western (continuum) knot of 32$\pm$4~mas 
based on five 
epochs of F606W images of the OT superimposed on the host galaxy. 

As seen from Fig.~\ref{host} the western and eastern knots seem to have
different broad band colours with the western component being the
reddest. To better quantify this we also retrieved the F814W HST
image from May 20 (lower panel in Fig.~\ref{hsthost} -- not drizzled due
to the lower S/N of the host in the F814W image) and measured the 
magnitudes of the two brightest
components of the host in F606W and F814W images. In these images
we measure the AB colours (V$_{606}$ $-$ I$_{814}$)$_{AB}$ = 
$-$0.11$\pm$0.11 and (V$_{606}$ $-$ I$_{814}$)$_{AB}$ = 0.19$\pm$0.07 for 
the eastern and western components respectively. Using the relation
between (V$_{606}$ $-$ I$_{814}$)$_{AB}$, redshift and the spectral 
slope $\beta$ ($f_{\lambda} \propto \lambda^{\beta}$)
derived by Meurer et al. (1999) we derive spectral slopes $\beta$ =
$-$2.4$\pm$0.3 and $-$1.4$\pm$0.2 for the two components. Based on the 
relation $A_{1600} = 4.43 + 1.99\beta$ also derived by Meurer et al.
(1999) we conclude that the eastern component must be essentially dust 
free (the derived value of $A_{1600}$ is formally negative, but consistent
with 0), whereas for the western component we find 
$A_{1600}$ = 1.6$\pm$0.4. The luminosity-weighted mean value of 
$A_{1600}$ for the sample of Lyman-Break galaxies studied by Meurer et 
al. (1999) is 1.8$\pm$0.2.

The restframe EW of 71$^{+20}_{-15}$ \AA \ is within the range of
50--200 \AA \ expected for dust-free galaxies (Charlot \& Fall 1993). 
Furthermore, the SFR derived from the Ly$\alpha$ flux in the 4\farcs0 
aperture taking into account the aperture correction is 6 M$_{\sun}$ 
yr$^{-1}$. This is only about a factor of 2 lower than what we found based 
on the restframe 2100 \AA \ continuum flux (Fynbo et al. 2001a). This means 
that there cannot be large amounts of dust in the host galaxy to absorb the 
resonantly scattered Ly$\alpha$ photons. This is consistent with the low
values of $A_{1600}$ derived above and with the
low SMC-like extinction measured from the optical afterglow (Fynbo et
al. 2001b; Price et al. 2001).

Of all the candidate LEGOs, the Ly$\alpha$ emission from the host 
galaxy of 
GRB~000926 is the only to have an obvious multi-component 
morphology. This suggests that we see several galaxy sub-clumps (or LEGOs 
as we call them) in the process of merging and that the GRB could be related 
to a merging induced starburst.

If we assume a R-band magnitude of R=28 for the GRB~000301C host galaxy
it is about 40 times fainter than GRB~000926 host galaxy. If the
GRB~000301C host has the same colour and Ly$\alpha$ equivalent width,
its Ly$\alpha$ flux will be $\sim$4$\times$10$^{-18}$ erg s$^{-1}$
cm$^{-2}$ which is a factor of 5 below the (5$\sigma$) detection limit 
in the narrow-band image. Hence, our data does not exclude that the 
GRB~000301C host galaxy is also a high equivalent width Ly$\alpha$ 
emitter.

\subsection{Exclusion of contamination due to foreground emission line 
sources}
For searches
for high redshift emission-line galaxies it is important
to realize that lower redshift galaxies with other emission
lines in the narrow filter can mimic high redshift Ly$\alpha$
candidates (e.g. Stern et al. 2000; Fynbo et al.
2001c). For the host galaxy of GRB~000926 the redshift is known from the
spectroscopy of the Optical Transient (OT) (Fynbo et al. 2001d; Castro et
al. 2001; M\o ller et al. in prep.). For the remaining candidate LEGOs 
we here consider other possible emission line sources than 
Ly$\alpha$ at z=2.04. We can exclude foreground galaxies with 
[\ion{O}{ii}] in the narrow filter, since such objects would have to 
be at redshift z=0 and hence be much brighter than our candidates. 
Distant planetary nebulae in the galaxy 
could be selected due to their strong \ion{O}{ii} emission lines, but such 
planetary nebulae are very rare at the high galactic latitudes of our 
fields (+44.4$^o$ and +37.3$^o$ respectively). In total only about 10 
planetary nebulae are known in the Milky Way halo (Howard et al. 1997).
Other possible explanations are \ion{C}{iv} 1550 \AA \ at z=1.39 or 
\ion{Mg}{ii} 2800 \AA \ at z=0.32. 
The presence of strong \ion{C}{iv} and \ion{Mg}{ii} emission requires 
an underlying AGN. The surface density of AGN at z$<$2.2 down to a 
B-band magnitude of 22.5 is 130 deg$^{-2}$ (Hartwick \& Schade 1990).
If we extrapolate to B=25.0 using the observed faint end slope of $d\log
N(B)/dB = 0.35$
we find 800 deg$^{-2}$. Hence, we expect 6 z$<$2.2 AGNs to this limit in 
each of our 27.6 arcmin$^2$ fields. The probability that these fall in the 
small volumes at z=0.32 and z=1.39 (0.18\% and 1.9\% of the total comoving 
volume out to z=2.2 respectively) probed by our filter is small,
especially since the space density of AGNs is declining rapidly with 
decreasing redshift. We therefore do not expect a 
significant \ion{C}{iv} and \ion{Mg}{ii} contamination due to AGN.
In conclusion, the only probable explanation for the excess emission in
the narrow filter is Ly$\alpha$ at z=2.04.

\section{Discussion and conclusions}
\label{discuss}

\subsection{LEGOs}

We first compare the photometric and other properties of the candidate 
LEGOs in the fields of GRB~000301C and GRB~000926 to properties 
of z$\approx$2 galaxies found in other surveys. 

Fontana et
al. (2000) have determined photometric redshifts for galaxies in the
HDF and NTT deep fields. From their catalogue we find 22 galaxies with 
photometric redshifts within $\Delta$z=0.15 of z=2.04. The I(AB)
magnitudes of these galaxies range from I(AB)=23.7 to I(AB)=25.8.
The range of I(AB) magnitudes for the candidate LEGOs in our fields is 
extending to somewhat fainter magnitudes as 4 of the 19 candidates
are not detected above the 2$\sigma$ level of 25.6 in the 3\farcs0 
aperture.

Kurk et al. (2000) and Pentericci et al. (2000) have detected
and spectroscopically confirmed 14 LEGOs in the
field around the radio galaxy PKS1138-262 using narrow-band
observations. They reach a 5$\sigma$ detection limit of
1.4$\times$10$^{-17}$ erg s$^{-1}$ cm$^{-2}$ over a field of 
38 arcmin$^2$ at z=2.156. Their filter spans a redshift range of
$\Delta z$ = 0.054. Their survey parameters are therefore 
similar to ours. The inferred number of LEGOs per arcmin$^2$
per unit redshift in their survey is 6.8$\pm$1.8. We expect to confirm
100\% (75\%) of our rank 1 (rank 2) candidates, hence 17 LEGOs
total in two 27.6 arcmin$^2$ fields observed with a 
filter spanning $\Delta z$ = 0.037. This corresponds to 8.3$\pm$2.0, 
which is consistent with the density found by Pentericci et al. 
(2000). 

Pentericci et al. (2000) estimate that their field is a factor of
6 overdense relative to the field, but this estimate is very
uncertain as it is based on {\it i)} only the four brightest objects in 
their sample, {\it ii)} a comparison with the spike of Lyman-Break 
galaxies observed by Steidel et al. (2000) at a significantly larger 
redshift (z=3.09) and with different selection criteria, and
{\it iii)} an assumption about the (unknown) fraction of galaxies at
z=2.16 that are Ly$\alpha$ emitters. Hence, we consider the evidence
for an overdensity to be at best weak. Unfortunately, there are currently 
no similar (in redshift and depth) 
surveys for Ly$\alpha$ emitters in blank fields so we  do not know yet
whether the QSO and GRB fields are both equally overdense 
or if they represent the mean density of LEGOs at z$\approx$2
(which we consider more likely). We are currently undertaking 
a large blank field survey at the Nordic Optical Telescope using the same 
filter with the purpose of examining the density and spatial distribution of
LEGOs in blank fields.

\subsection{GRB host galaxies}

This study confirms what is already known from broad-band studies,
namely that GRBs occur in galaxies covering a very broad range of
luminosities. The faintest host galaxies detected so far in the
broad-bands are the hosts of GRB~000301C and GRB~990510 at
R$\approx$29.7 and R$\approx$29.2 and the host of GRB~980326 at
V$\approx$30.3 (Levan, private communication; Fruchter et al. in prep.).

We have used deep Ly$\alpha$ narrow band imaging to map out the
star formation in volumes around the host galaxies of two GRBs.
GRB~000926 occurred in one of the strongest centres of star formation
within several Mpc, whereas GRB~000301C occurred in an intrinsically
very faint galaxy far from being the strongest centre of star formation
in its galactic environment (cf. Table~\ref{candprop1}).

What is the underlying mechanism for GRBs? The answer to this
question is still not known, but we now have several vital clues.
It has been noted that GRB host galaxies generally are UV bright
and have emission
lines with large equivalent widths (Bloom et al. 2002; Fruchter et al.
in prep.), indicating that they are actively star forming. This result
is supported by our detection of strong Ly$\alpha$ emission from the
host of GRB~000926, and is not contradicted by the lack of Ly$\alpha$
emission from the host of GRB~000301C where even a very high Ly$\alpha$
rest equivalent width of $\sim$150 \AA \ would have remained undetectable
in our deep narrow band image. It is quite natural therefore, to
hypothesize that the GRB phenomenon is linked to star formation
(Paczy{\'n}ski 1998).
A strong prediction from this hypothesis is that the GRB host galaxies
are drawn from the underlying population of high redshift galaxies
weighted with the integrated star formation rate {\it in each magnitude
bin}. If the LF of the underlying population was flat, then we would
predict that all GRB host galaxies should be very bright.
Conversely a very steep LF would result in all GRB host galaxies being
drawn from the very faintest galaxies. The observation that GRB host
galaxies span a very wide range of luminosities therefore places
constraints on the allowed range of faint end slopes of the
LF. For example, if we use a standard Schechter form
\[
\Phi(L)dL = \Phi^* \left(\frac{L}{L^*}\right)^{\alpha} \exp \left(-
\frac{L}{L^*}\right)d\left(\frac{L}{L^*}\right)
\]
of the LF, and use R$^*$=23.5\footnote{For Lyman-Break galaxies at
z=3.0 Adelberger \& Steidel (2000) find R$^*$=24.54, which roughly
corresponds to R=23.5 at z=2.0}
then a simple calculation shows that the observed flat
distribution in the range R=24 to R=30 requires a faint end slope
$\alpha=-1.9\pm 0.7$. Such a steep (steeper than in the local
universe) faint end slope is in good agreement with the faint end
slope for Lyman-Break galaxies at z=3.0 ($\alpha=-1.57\pm0.11$,
Adelberger \& Steidel, 2000), and with the faintness and small impact
parameters of high redshift DLA galaxies (Fynbo et al. 1999; M{\o}ller
et al. 2001). The current data set is therefore compatible with the
hypothesis that GRBs trace star formation, and the outlook for
impending advance on this question is good. A full statistical treatment
of a slightly larger sample of GRB host galaxies would significantly
decrease the error bars on $\alpha$, and would show if it
remains compatible with the faint end slope of the LBGs. More
importantly, the GRB host galaxies probe fainter magnitudes than the
LBG samples and will therefore reveal if there is a turn-over, or if
the faint end slope continues. The question of how far the faint end
of the LF continues with unchanged slope is important for the
determination of the total star formation rates at high redshifts.

\section*{Acknowledgements}
The data presented here have been taken using ALFOSC, which is owned
by the Instituto de Astrofisica de Andalucia (IAA) and operated at the
Nordic Optical Telescope under agreement between IAA and the NBIfAFG
of the Astronomical Observatory of Copenhagen. MA acknowledges the 
support of the U. of Oulu astrophysics group. JG acknowledges the
the receipt of  a Marie  Curie  Research Grant from  the European
Commission. MPE and MW acknowledge support from the ESO Directors 
Discretionary Fund. JUF, MPE and MW acknowledges excellent support 
at the NOT during the 5 nights in May at which most of the data 
presented here were obtained. This work was supported by the Danish 
Natural Science Research Council (SNF). JUF is supported by an ESO 
research fellowship. We thank P. Vreeswijk and R. Mendez 
for helpful discussions and our referee N. Panagia for
comments that improved our manuscript on several important points.

\appendix
\section{Errors on colours for faint sources}
Here we detail a method by which to calculate two-sided error-bars 
for sources detected at low signal-to-noise ratio. We wish to calculate
the colour $m(a)-m(b)$ for a source with fluxes $a$ and $b$ in two
passbands. We first derive the likelihood function for the flux ratio 
$\alpha = b/a$. To do this we introduce $\beta = a \cdot b$ and note that 
$b^2 = \alpha \cdot \beta$ and $a^2 = \beta / \alpha$. The joint likelihood
function for $\alpha$ and $\beta$ can then be found as

\begin{tiny}
\[
L(\alpha,\beta) \propto \exp{( -\frac{(\sqrt{\beta/\alpha}-\sqrt{\beta_0/\alpha_0})^2}{2\sigma_a^2}
- \frac{(\sqrt{\alpha \cdot \beta} - \sqrt{\alpha_0 \cdot \beta_0})^2}{2\sigma_b^2})},
\]
\end{tiny}

\noindent
where ($\alpha_0,\beta_0$) is the maximum likelihood estimate.
By maximizing the likelihood function in the variable $\beta$ for a
fixed value of $\alpha$ we find 
\[
\sqrt{\frac{\beta}{\alpha}} = 
\frac{ \frac{b_0}{\sigma_b^2} + \frac{a_0}{\sigma_a^2 \alpha}  }
{ \frac{1}{\alpha \sigma_a^2} + \frac{\alpha}{\sigma_b^2} } 
\]
\noindent
and
\[
\sqrt{\alpha \cdot \beta } = 
\frac{ \frac{b_0 \alpha}{\sigma_b^2} + \frac{a_0}{\sigma_a^2}  }
{ \frac{1}{\alpha \sigma_a^2} + \frac{\alpha}{\sigma_b^2} } 
\]
\noindent
After inserting these expressions we find the likelihood function for
$\alpha$ of the form
\begin{small}
\[
\log L(\alpha) = \log L(\alpha_0) - \frac{1}{2}
\frac{a_0^2\sigma_a^2 (\alpha-\alpha_0)^2 + b_0^2\sigma_b^2
(1/\alpha-1/\alpha_0)^2} {\sigma_b^2 \alpha^{-1} + \sigma_a^2 \alpha}
\]
\end{small}
\noindent
We then determine the two-sided error-bars for the ratio $\alpha$ as
the two solutions to the equation 
\[
\log L(\alpha_0+\sigma_{\alpha}) = \log L(\alpha_0) - 1/2
\]
\noindent
Finally, we use the error-bars on $\alpha$ to calculate the error-bars
on the colour in the same way as for the magnitudes in Sect.~\ref{photometry}.

\end{document}